\documentstyle[12pt]{article}
\begin{document}


\renewcommand{\thesection}{\arabic{section}}
\renewcommand{\theequation}{\arabic{equation}}
\renewcommand {\c}  {\'{c}}
\newcommand {\cc} {\v{c}}
\newcommand {\s}  {\v{s}}
\newcommand {\CC} {\v{C}}
\newcommand {\C}  {\'{C}}
\newcommand {\Z}  {\v{Z}}
\newcommand{\pv}[1]{{-  \hspace {-4.0mm} #1}}

\baselineskip=24pt

\def\beqra{\begin{eqnarray}} \def\eeqra{\end{eqnarray}}
\def\beqast{\begin{eqnarray*}} \def\eeqast{\end{eqnarray*}}
\def\beq{\begin{equation}}      \def\eeq{\end{equation}}
\def\be{\begin{enumerate}}   \def\ee{\end{enumerate}}

\def\gam{\gamma}
\def\Gam{\Gamma}
\def\la{\lambda}
\def\eps{\epsilon}
\def\La{\Lambda}
\def\si{{\rm si}}
\def\Si{\Sigma}
\def\al{\alpha}
\def\Th{\Theta}
\def\th{\theta}
\def\tnu{\tilde\nu}
\def\vphi{\varphi}
\def\del{\delta}
\def\Del{\Delta}
\def\ab{\alpha\beta}
\def\om{\omega}
\def\Om{\Omega}
\def\mn{\mu\nu}
\def\mun{^{\mu}{}_{\nu}}
\def\kap{\kappa}
\def\rsi{\rho\sigma}
\def\beal{\beta\alpha}
\def\til{\tilde}
\def\rta{\rightarrow}
\def\eqv{\equiv}
\def\nab{\nabla}
\def\pa{\partial}
\def\sit{\tilde\sigma}
\def\ul{\underline}
\def\indt{\parindent2.5em}
\def\nd{\noindent}
\def\rsi{\rho\sigma}
\def\beal{\beta\alpha}
\def\caa{{\cal A}}
\def\cb{{\cal B}}
\def\cac{{\cal C}}
\def\cd{{\cal D}}
\def\ce{{\cal E}}
\def\cf{{\cal F}}
\def\cg{{\cal G}}
\def\cah{{\cal H}}
\def\ci{{\rm ci}}
\def\cj{{\cal{J}}}
\def\ck{{\cal K}}
\def\cl{{\cal L}}
\def\cm{{\cal M}}
\def\cn{{\cal N}}
\def\cO{{\cal O}}
\def\cp{{\cal P}}
\def\car{{\cal R}}
\def\cs{{\cal S}}
\def\ct{{\cal{T}}}
\def\cu{{\cal{U}}}
\def\cv{{\cal{V}}}
\def\cw{{\cal{W}}}
\def\cx{{\cal{X}}}
\def\cy{{\cal{Y}}}
\def\cz{{\cal{Z}}}
\def\asymptotic{{_{\stackrel{\displaystyle\longrightarrow}
{x\rightarrow\pm\infty}}\,\, }} 
\def\asymptext{\raisebox{.6ex}{${_{\stackrel{\displaystyle\longrightarrow}
{x\rightarrow\pm\infty}}\,\, }$}} 
\def\epsilim{{_{\textstyle{\rm lim}}\atop
_{~~~\epsilon\rightarrow 0+}\,\, }} 
\def\omegalim{{_{\textstyle{\rm lim}}\atop
_{~~~\om^2\rightarrow 0+}\,\, }} 
\def\xlimp{{_{\textstyle{\rm lim}}\atop
_{~~x\rightarrow \infty}\,\, }} 
\def\xlimm{{_{\textstyle{\rm lim}}\atop
_{~~~x\rightarrow -\infty}\,\, }} 
\def\asymptoticp{{_{\stackrel{\displaystyle\longrightarrow}
{x\rightarrow +\infty}}\,\, }} 
\def\asymptoticm{{_{\stackrel{\displaystyle\longrightarrow}
{x\rightarrow -\infty}}\,\, }} 

\def\raisenot{\raise .5mm\hbox{/}}
\def\nota{\ \hbox{{$a$}\kern-.49em\hbox{/}}}
\def\notA{\hbox{{$A$}\kern-.54em\hbox{\raisenot}}}
\def\notb{\ \hbox{{$b$}\kern-.47em\hbox{/}}}
\def\notB{\ \hbox{{$B$}\kern-.60em\hbox{\raisenot}}}
\def\notc{\ \hbox{{$c$}\kern-.45em\hbox{/}}}
\def\notd{\ \hbox{{$d$}\kern-.53em\hbox{/}}}
\def\notbd{\ \hbox{{$D$}\kern-.61em\hbox{\raisenot}}} 
\def\note{\ \hbox{{$e$}\kern-.47em\hbox{/}}}
\def\notk{\ \hbox{{$k$}\kern-.51em\hbox{/}}}
\def\notp{\ \hbox{{$p$}\kern-.43em\hbox{/}}}
\def\notq{\ \hbox{{$q$}\kern-.47em\hbox{/}}}
\def\notW{\ \hbox{{$W$}\kern-.75em\hbox{\raisenot}}}
\def\notz{\ \hbox{{$Z$}\kern-.61em\hbox{\raisenot}}}
\def\notpa{\hbox{{$\partial$}\kern-.54em\hbox{\raisenot}}}
\def\fo{\hbox{{1}\kern-.25em\hbox{l}}}  
\def\rf#1{$^{#1}$}
\def\bx{\Box}
\def\tr{{\rm Tr}}
\def\rmtr{{\rm tr}}
\def\dgg{\dagger}
\def\lag{\langle}
\def\rag{\rangle}
\def\bmid{\big|}
\def\vlap{\overrightarrow{\La p}} 
\def\lrta{\longrightarrow} \def\lrar{\raisebox{.8ex}{$\longrightarrow$}}
\def\ON{{\cal O}(N)}
\def\UN{{\cal U}(N)}
\def\bdPh{\mbox{\boldmath{$\dot{\!\Phi}$}}}
\def\bPh{\mbox{\boldmath{$\Phi$}}}
\def\bPhs{\bPh^2}
\def\sef{S_{eff}[\sigma,\pi]}
\def\sigx{\sigma(x)}
\def\pix{\pi(x)}
\def\bph{\mbox{\boldmath{$\phi$}}}
\def\bphs{\bph^2}
\def\ex{\BM{x}}
\def\exs{\ex^2}
\def\xdot{\dot{\!\ex}}
\def\y{\BM{y}}
\def\ys{\y^2}
\def\ydot{\dot{\!\y}}
\def\pat{\pa_t}
\def\pax{\pa_x}
\def\hp{{\pi\over 2}}
\def\sign{{\rm sign}\,}
\def\bv{{\bf v}}




\begin{center}
{\bf  Collective Field Formulation of the Multispecies Calogero Model and its 
Duality Symmetries}
\bigskip

V. Bardek$^{a}${ \footnote{e-mail: bardek@irb.hr}},
J. Feinberg$^{b,c}${ \footnote{e-mail: joshua@physics.technion.ac.il}} 
 \hspace{0.2cm} and \hspace{0.2cm}
S.Meljanac$^{a}$ {\footnote{e-mail: meljanac@irb.hr}} \\  
$^{a}$ Rudjer Bo\v{s}kovi\'c Institute, Bijeni\v cka  c.54, HR-10002 Zagreb,
Croatia \\[3mm] 

$^{b}$  Department of Physics, University of Haifa at Oranim, Tivon 36006, 
Israel,\newline and \\
$^{c}$ Department of Physics, Technion-Israel Institute of Technology, 
Haifa 32000, Israel \\[3mm]

\bigskip

\end{center}
\setcounter{page}{1}
\bigskip


\begin{minipage}{5.8in}

{\abstract~~~We study the collective field formulation of a restricted form 
of the multispecies Calogero model, in which the three-body interactions are 
set to zero. We show that the resulting collective field theory is invariant 
under certain duality transformations, which interchange, 
among other things, particles and antiparticles, and thus generalize the 
well-known strong-weak coupling duality symmetry of the ordinary Calogero 
model. We identify all these dualities, which form an Abelian group, and study 
their consequences. We also study the ground state and small 
fluctuations around it in detail, starting with the two-species model, and 
then generalizing to an arbitrary number of species.}
 
\end{minipage}

\bigskip
PACS number(s): 03.65.Sq, 05.45.Yv, 11.10.Kk, 11.15.Pg 


\newpage


\section{Introduction}
The Calogero Model (CM) \cite{Calogero:1969xj} - \cite{ Brink:1993sz}
 is a well-known exactly solvable many-body 
system, both at the classical and quantum levels. The CM and its various 
descendants continue to draw considerable interest due to their diverse
physical applications in systems such as random matrices \cite{lee}, 
fractional statistics \cite{Haldane:1987gg}-\cite{ mms},
 gravity and black hole physics \cite{Gibbons:1998fa, Birmingham:2001qa}, spin chains 
\cite{SriramShastry:1987gh, Haldane:1991xg},
 solitons \cite{Polychronakos:1994xg}-\cite{ abanov},
 2D Yang-Mills theory \cite{Gorsky:1993pe, Minahan:1993mv}, lowest Landau level (LLL) anyon 
models \cite{DasnieresdeVeigy:1993rx, isakov},
 Chern-Simons matrix model \cite{Susskind:2001fb}-\cite{ hellerman},
 Laughlin-Hall states \cite{Jonke:2001cm}-\cite{ ms}, and 
unoriented superstrings in two dimensions \cite{Gomis:2003vi}.

Calogero's original model describes $N$ indistinguishable particles on the 
line which interact through an inverse-square two-body interaction. 
It is well-known, however, that the CM may alternatively be interpreted in 
terms of $N$ free particles obeying generalized exclusion statistics
 \cite{Polychronakos:1999sx, Polychronakos:2006, Murthy:1994gk, wu}.

Haldane`s formulation of statistics may be extended to systems made of 
different species of particles, in which the interspecies statistical
coupling depends on the species being coupled. This may be implemented 
in a multi-species generalization of the CM in which particles have 
different masses and different couplings to each other
\cite{forrester}-\cite{ mask}.

Quite a few such generalized multi-sepcies Calogero models exist, but 
contrary to the original CM, knowledge about their exact solvability was 
rather tenuous. The recent breakthrough in this front derives from the 
papers \cite{sergeev1}-\cite{ Guhr:2004ff}. The authors of \cite{sergeev1}
introduced deformed Calogero models, 
related to root systems of superalgebras, and gave effectively
a proof of their integrability. In \cite{sergeev2} they presented a more conceptual 
proof by using shifted super-Jack polynomials. In related developments, 
the authors of \cite{guhr, Guhr:2004ff} introduced a supersymmetric generalizations of the CM 
which was based on Jacobians for the radial coordinates on certain 
superspaces. Both aforementioned models are closely related to the 
multi-family generalization of the CM introduced in
\cite{Meljanac:2003jj, Meljanacstojic}.

Motivated by these developments, in the present paper we investigate
the latter model in the limit in which each family contains a large number of 
particles. In this limit, the high-density limit, the system is amenable to 
large-$N$ collective-field formulation. As is well-known, the collective
theory offers a continuum field-theoretic framework for studying interesting
aspects of many-particle systems, somwhat analogous to the continuum 
hydrodynamic description of fluids. It is appropriate to mention at this 
point the recent review on the collective-field 
and other continuum approaches to the spin-Calogero-Sutherland model
\cite{Aniceto-Jevicki}. 
The collective formulation has several virtues. In the large-$N$ limit 
dynamics is governed by saddle points of the effective collective action,
which contains the leading quantum effects. Thus, by extremizing this action, 
one is able to compute the uniform ground-state collective-field 
configuration, as well as topological and non-topological soliton 
configurations, and their corresponding energy eigenvalues. 
By expanding around these extrema it is possible to go beyond the 
large-$N$ leading order and obtain the spectrum and wave-functionals of the 
quadratice fluctuations around these semiclassical configurations, 
and also to compute the corresponding density-density correlation functions. 

Beside these obvious advantages, the collective-field theory provides a 
natural framework for analyzing symmetries of the system which cannot be seen 
directly in the original (finite) $N$-particle quantum system. An important 
example in this respect is the strong-weak coupling duality symmetry of the 
one-family Calogero model discussed in \cite{Minahan:1994ce}. 
In this paper we generalize this approach to the multi-family Calogero model. 
We show that the resulting collective field theory is invariant 
under certain duality transformations, which interchange, 
among other things, particles and antiparticles, and thus generalize the 
duality symmetry \cite{Minahan:1994ce} of the ordinary Calogero 
model. We identify all these dualities, which form an Abelian group, and study 
their consequences. In particular, the investigations carried in this 
paper will enable us to find the conditions under which collective 
quasi-particles describing density fluctuations in the the $F$-family 
Calogero model can be identified with those of an effective one-family 
Calogero model. 
As a by-product, this may help to better understand the exact solvability 
of some of the recently proposed two-family Calogero models 
\cite{sergeev1}-\cite{ Guhr:2004ff}. We stress that the duality relations 
derived and discussed in this paper are {\em exact} symmetries of the 
collective-field hamiltonian, as opposed to the approxiamte duality symmetries
discussed in \cite{andrjur, andrjur1}\,.

This paper is organized as follows: In Section 2 we review some 
known facts about the two-family Calogero model \cite{Bardek:2005yx}. 
We then derive the quantum collective-field Hamiltonian, including 
divergent terms which are inherent to the collective-field formulation. 
In Section 3 we show that this collective field theory is invariant 
under certain duality transformations, which interchange, 
among other things, particles and antiparticles, and thus generalize the 
well-known strong-weak coupling duality symmetry of the ordinary Calogero 
model. We make a complete list of these dualities, show that they form 
an Abelian group, and study their consequences.
In Section 4 we concentrate on the leading large-$N$ behavior of the 
collective Hamiltonian of the two-family model and identify the 
effective potential which dominates dynamics in this limit. 
The large-$N$ uniform ground state configuration is then found by minimizing 
the effective potential, and the fluctuations around it are then investigated.
The Hamiltonian which governs these small quadratic density fluctuations 
contains divergent terms. By carefully choosing the creation and annihilation 
operators describing these excitations, these divergent terms can be renormalized away. 
Section 5 is devoted to diagonalizing the latter Hamiltonian, resulting in two 
{\em decoupled} quasi-particle dispersions. One of these dispersions, at 
small momenta, has the structure of a single-species Calogero model with 
some effective particle density, mass and statistical parameter, which we 
calculate. The other dispersion is that of free massive particles. 
In Section 6 we consider the ground-state wave-functional of small 
fluctuations, and calculate the corresponding static density-density 
correlation functions of the two-family model. Not surprisingly, the 
long-distance behavior of these correlation functions coincides with that of 
the effective single-species Calogero model alluded to above. 
In Section 7, we generalize our study of the two-family case to the 
$F$-family Calogero model (with no three-body interactions). We find all 
the exact duality transformations which leave the collective Hamiltonian 
of the $F$-family Calogero model invariant, and identify the 
Abelian group they form. We also obtain the ground-state and study 
the small fluctuations around it. In particular, we compute the 
low-momentum behavior of the dispersion relations of the quasi-particles 
corresponding to these fluctuations. As in the two-family case, one of these 
dispersions has the structure of a single-species Calogero model with 
some effective particle density, mass and statistical parameter. The 
remaining $F-1$ dispersions are those of free massive particles.
We then compare our dispersion laws with those found before in the 
generalized Thomas-Fermi approach \cite{Sen:1995zj}. 
Finally, in Section 8 we argue how our collective-field theory 
approach may explain the exact solvability of the two-family Calogero models 
proposed in recent papers.


\section{Two-family collective-field Hamiltonian}

In order to get oriented, and in preparation for studying the multi-species 
Calogero model, let us start by recalling the essential features of the 
two-family Calogero system. The Hamiltonian of this model reads \cite{ Meljanacstojic}
$$
H = - \frac{1}{2 m_{1}} \sum_{i=1}^{N_{1}} 
\frac{{\partial}^{2}}{\partial {x_{i}}^{2}} +
\frac{\lambda_{1} (\lambda_{1} - 1)}{2 m_{1}} \sum_{i \neq j
}^{N_{1}} \frac{1}{{(x_{i} - x_{j})}^{2}} -  \frac{1}{2 m_{2}}
\sum_{\alpha = 1}^{N_{2}} \frac{{\partial}^{2}}{\partial {x_{\alpha}}^{2}} +
\frac{\lambda_{2} (\lambda_{2} - 1)}{2 m_{2}} \sum_{\alpha \neq \beta }^{N_{2}}
\frac{1}{{(x_{\alpha} - x_{\beta})}^{2}}  
$$
\begin{equation} \label{h1}
 + \frac{1}{2} \left( \frac{1}{ m_1}  +  \frac{1} { m_2 } \right)
 \lambda_{12}(\lambda_{12} -1)
 \sum_{i = 1}^{N_{1}}\sum_{\alpha = 1 }^{N_{2}}
 \frac{1}{(x_{i}-x_{\alpha})^{2}}.
\end{equation}
Here, the first family contains $ \; N_{1} \; $ particles of mass 
$ \; m_{1} \; $ at positions $ \; x_{i}, \; i = 1,2,...,N_{1}, \; $ and 
the second one contains $ \; N_{2} \; $ particles of mass $ \; m_{2} \; $ 
at positions $ \; x_{\alpha}, \; \alpha = 1,2,...,N_{2}. $ 
All particles interact via two-body inverse-square potentials. 
The interaction strengths within each family are parametrized by the coupling
constants $ \; \lambda_{1} \; $ and $ \; \lambda_{2}, \; $ respectively.  
The interaction strength between particles of the first and the second family
is parametrized by $ \; \lambda_{12}.$ For reasons that should become clear in 
later sections, we shall assume in this paper that $\lambda_1$ and $\lambda_2$ 
are always positive, whereas the sign of $\lambda_{12}$ is unrestricted. In any case, 
these couplings are always assumed to 
be set such that (\ref{h1}) has a well-defined ground state. 

Note that we did not include in (\ref{h1}) a confining potential. This is not
really a problem, as we can always add a very shallow confining potential to 
regulate the problem (in the case of purely repulsive interactions), or else, 
consider the particles confined to a very large circle (i.e., consider
(\ref{h1}) as the large radius limit of the Calogero-Sutherland model \cite{Sutherland:1971ep}). 
We shall henceforth tacitly assume that the system is thus properly 
regularized at large distances.

In (\ref{h1}) we imposed the restriction that there be no three-body 
interactions, which requires \cite{Meljanac:2003jj, Meljanacstojic,  Meljanac:2004mr, 
Meljanac:2004vi, Meljsams}
\begin{equation} \label{threebody}
  \frac{\lambda_{1}}{{m_{1}}^{2}} = \frac{\lambda_{2}}{{m_{2}}^{2}} = 
 \frac{\lambda_{12}}{m_{1} m_{2}}.
\end{equation} 
This particular relation was also displayed in \cite{Sen:1995zj}, as a consequence
of the requirement that the asymptotic Bethe Ansatz should be applicable to
the ground state of the multi-species model.
It follows from (\ref{threebody}) that 
\beq\label{lambda12}
\lambda_{12}^2 = \lambda_1\lambda_2\,.
\eeq
We assume that (\ref{lambda12}) holds throughout this paper. 
The Hamiltonian (\ref{h1}) describes the simplest multi-species Calogero 
model for particles on the line, interacting only with 
two-body potentials. The singularities of the Hamiltonian (\ref{h1}) at 
points where particles coincide implies that the many-body eigenfunctions 
contain a product of Jastrow-type prefactors $\Pi_1\Pi_2\Pi_{12}$, where
\beqra\label{jastrow}
\Pi_{1} &=& \prod_{i < j}^{N_{1}} { (x_{i} - x_{j})}^{\lambda_{1}}\,,
\nonumber\\
\Pi_{2} &=& \prod_{\alpha < \beta}^{N_{2}} {( x_{\alpha} - x_{\beta})}^{\lambda_{2}}\,,\nonumber\\
\Pi_{12} &=& \prod_{i, \alpha}^{N_{1},N_{2}} {( x_{i} - x_{\alpha})}^{\lambda_{12}}\,.
\eeqra
These Jastrow factors vanish (for positive $\lambda$'s) at particle coincidence points, and 
multiply that part of the wave-function which is totally symmetric under any permutation
of {\em identical} particles. It is precisely these symmetric wave-functions
on which the collective field operators act, as explained below. 

Let us make a couple of elementary comments at this point: 
First, we note from (\ref{jastrow}) that for $ \; \lambda_{i}=0 \; $ and 
$ \;\lambda_{i}=1, \; $ the model describes two families of interacting 
bosons and fermions, respectively.
Second, superficial glance at (\ref{h1}) may suggest that the model breaks 
into two independent Calogero systems at $ \; \lambda_{12}=1 \; $. This is 
wrong, however. The two families remain correlated at 
$ \; \lambda_{12}=1, \; $ and display mutual fermionic behavior. 
Namely, the relevant prefactor in (\ref{jastrow}) picks up a factor 
$ \; {(-1)}^{ \lambda_{12}} \; $ under the exchange of particle indices 
$ \; ( i \leftrightarrow \alpha). $ We will see, after making the 
trasformation to collective variables, that this gives a non-trivial 
interacting bosonic theory of the two collective fields describing different 
families of particles.

Let us recall at this point some of the basic ideas of the 
collective-field method (adapted for our two-family Calogero model \cite{Bardek:2005yx}): 
Instead of solving the Schr\"odinger equation associated with (\ref{h1}) 
for the many-body eigenfunctions, subjected to the appropriate particle 
statistics (Bosonic, Fermionic of fractional), we restrict ourselves to 
functions which are totally symmetric under any permutation of identical 
particles. This we achieve by stripping off the Jastrow factors 
(\ref{jastrow}) from the eigenfunctions, which means performing on 
(\ref{h1}) the similarity transformation
\begin{equation} \label{similaritytr}
 H \rightarrow \tilde H = \Pi_{12}^{-1} \Pi_{2}^{-1} \Pi_{1}^{-1} H \Pi_{1} 
\Pi_{2} \Pi_{12}\,,
\end{equation}
where the Hamiltonian
$$
 \tilde H = - \frac{1}{2 m_{1}} \sum_{i=1}^{N_{1}} \frac{{\partial}^{2}}
{\partial {x_{i}}^{2}} -
    \frac{1}{m_{1}} \left( \lambda_{1} \sum_{i \neq j}^{N_{1}}
   \frac{1}{x_{i} - x_{j}} + \lambda_{12} \sum_{i, \alpha} \frac{1}{x_{i} - 
x_{\alpha}} \right) \frac{\partial}{\partial x_{i}}
$$
\begin{equation} \label{h1tr}
    - \frac{1}{2 m_{2}} \sum_{\alpha = 1}^{N_{2}} \frac{{\partial}^{2}}
{\partial {x_{\alpha}}^{2}} -
    \frac{1}{m_{2}} \left( \lambda_{2} \sum_{\alpha \neq \beta}^{N_{2}}
   \frac{1}{x_{\alpha} - x_{\beta}} + \lambda_{12} \sum_{i, \alpha} \frac{1}
{x_{\alpha} - x_{i}} \right)
        \frac{\partial}{\partial x_{\alpha}}\,.
\end{equation}
Note that $\tilde H$ does not contain the singular two-body interactions. 
By construction, this Hamiltonian is hermitian with respect to the measure 
$$d\mu (x_i, x_\alpha) = (\Pi_{1} \Pi_{2}\Pi_{12} )^2\, 
d^{N_1} x_i\, d^{N_2} x_\alpha\,,$$ 
(as opposed to the original Hamiltonian $H$ in (\ref{h1}), which is 
hermitian with respect to the flat Cartesian measure).

We can think of the symmetric many-body wave-functions acted upon by 
$\tilde H$ as functions depending on all possible symmetric
combinations of particle coordinates (symmetric separately in the coordinates
of each family of particles). These combinations form an overcomplete set of 
variables. However, as explained below, in the {\em continuum} limit, 
redundancy of these symmetric variables has a negligible effect. 
The set of these symmetric variables can be generated, for example, 
by producs of moments of the collective  - or density - fields 
\beq\label{collective}
\rho_{1}(x) = \sum_{i = 1}^{N_{1}} \delta( x - x_{i})\,,\quad\quad 
\rho_{2}(x) = \sum_{\alpha = 1}^{N_{2}} \delta( x - x_{\alpha})\,.
\eeq
The collective-field theory for the two-family Calogero model is obtained
by changing variables from the particle coordinates $ \; x_{i} \; $ and
$ \; x_{\alpha} \; $ to the density fields $ \; \rho_{1}(x) \; $ and
$ \; \rho_{2}(x) \; $. This transformation replaces the finitely many variables
$ \; x_{i} \; $ and $ \; x_{\alpha} \; $ by two continuous fields, which is 
just another manifestation of overcompleteness of the collective variables.  
Clearly, description of the particle systems in terms of continuous fields
becomes an effectively good description in the high density limit. 
In this limit the mean interparticle distance is much smaller than than any 
relevant physical length-scale, and the $\delta$- spikes in (\ref{collective})
can be smoothed out into well-behaved countinuum fields. All this is in direct
analogy to the hydrodynamical effective description of fluids, which replaces 
the microscopic atomistic formulation. Of course, the large density limit 
means that we have taken the large- $N_1, N_2$ limit. (It is 
understood throughout this paper that $N_1$ and $N_2$ tend to infinity at 
comparable rates.)

Changing variables from particle coordinates $x_i, x_\alpha$ to the 
collective fields (\ref{collective}) implies that we should express all 
partial derivatives in  the Hamiltonian $\tilde H$ in (\ref{h1tr}) as 
\beq\label{derivatives} 
\frac{\partial}{\partial x_{i}} = \int dx \frac{\partial \rho_{1}(x)}
{\partial x_{i}} \frac{\delta}{\delta \rho_{1}(x)}\,,\quad\quad
\frac{\partial}{\partial x_{\alpha}} = \int dx \frac{\partial \rho_{2}(x)}
{\partial x_{\alpha}}\frac{\delta}{\delta \rho_{2}(x)}\,,
\end{equation}
where we applied the differentiation chain rule. 

In the large $- N_1, N_2 $ limit, the Hamiltonian $\tilde H$ can be 
expressed entirely in terms of the collective fields $ \; \rho_{1},\; 
\rho_{2}\; $ and their canonical conjugate momenta 
\beq\label{momenta}
\pi_{1}(x) = - i \frac{\delta}{\delta \rho_{1}(x)}\,,\quad\quad
\pi_{2}(x) = - i \frac{\delta}{\delta \rho_{2}(x)}\,,
\end{equation}
as we show below. It follows from (\ref{derivatives}) and (\ref{momenta}) 
that the particle momentum operators (acting on symmetric wave-functions)
may be expressed in terms of the collective-field momenta at particular
points on the line as 
\beq\label{particlemomenta}
p_i = -\pi_1'(x_i)\,,\quad\quad p_\alpha = -\pi_2'(x_\alpha)\,,
\eeq
(where $\pi_a'(x) = \pax \pi_a (x)$). 
Finally, note from (\ref{collective}) that the collective fields obey the 
normalization conditions
\beq\label{conservation}
\int dx \rho_{1}(x) = N_{1}\,,\quad\quad \int dx \rho_{2}(x) = N_{2}\,.
\eeq

The density fields $ \; \rho_{1}, \rho_{2}, \; $ and their conjugate 
momenta $ \; \pi_{1}, \pi_{2}, \; $ satisfy the equal-time canonical 
commutation relations\setcounter{footnote}{0}\footnote{According to 
(\ref{conservation}), the zero-momentum modes of the density fields are 
constrained, i.e., non-dynamical. This affects the first set of commutation 
relations in (\ref{canonical}), whose precise form is $[\rho_a (x), \pi_b (y)]
= i \delta_{ab}(\delta(x - y) - (1/l))$, where $l$ is the size of the large 
one-dimensional box in which the system is quantized, which is much larger 
than the macroscopic size $L$ of the particle condensate in the system. 
In what follows, we can safely ignore this $1/l$ correction in the commutation
relations.}
\beq\label{canonical}
[\rho_a (x), \pi_b (y)] = i \delta_{ab}\delta(x - y)\,,\quad a,b=1,2\,,
\eeq
\beqast
[\rho_a (x), \rho_b (y)] = [\pi_a (x), \pi_b (y)] = 0\,.\quad\quad {}
\eeqast

By substituting (\ref{collective})-(\ref{particlemomenta}) in (\ref{h1tr}), 
we obtain the continuum-limit expression for $\tilde H$ as 
$$
\tilde H = \frac{1}{2 m_{1}} \int dx \rho_{1}(x) { ( \partial_{x} 
\pi_{1}(x))}^{2}
$$
$$ 
 - \frac{i}{m_{1}} \int dx \rho_{1}(x) \left( \frac{\lambda_{1} - 1}{2} 
\frac{\partial_{x} \rho_{1}}{\rho_{1}} +
\lambda_{1} \pv \int \frac{ dy \rho_{1}(y)}{x - y}  + \lambda_{12} \pv \int 
\frac{dy \rho_{2}(y)}{x - y} \right) \partial_{x} \pi_{1}(x)
$$
$$
 + \frac{1}{2 m_{2}} \int dx \rho_{2}(x){(\partial_{x} \pi_{2}(x))}^{2} 
$$
\begin{equation}\label{htilde}
 - \frac{i}{m_{2}} \int dx \rho_{2}(x) \left( \frac{\lambda_{2} - 1}{2} 
\frac{\partial_{x} \rho_{2}}{\rho_{2}}
  + \lambda_{2} \pv \int \frac{ dy  \rho_{2}(y)}{x - y}
  + \lambda_{12} \pv \int \frac{dy \rho_{1}(y)}{x - y} \right) \partial_{x} 
\pi_{2}(x),
\end{equation}
where $ \; \pv \int \; $ denotes Cauchy's principal value.

It can be shown \cite{sakita} that (\ref{htilde}) is hermitian with 
respect to the functional measure\footnote{By definition (recall 
(\ref{collective})), this measure is defined only over positive values of 
$\rho_1, \rho_2$\,.}
\beq\label{functionalmeasure} 
{\cal D} \mu [\rho_1, \rho_2] = J[\rho_1, \rho_2] \prod_x 
d\rho_1(x)\,d\rho_2(x)\,,
\eeq
where $ J[\rho_1, \rho_2] $ is the Jacobian of the transformation from 
$ \; \{ x_{i}, x_{\alpha} \} \; $ to the collective fields 
$ \; \{ \rho_{1}(x), \rho_{2}(x) \}$\,. In the large - $N_1, N_2$ limit 
it is given by \cite{Bardek:2005yx} 
$$
 \ln J = (1 - {\lambda}_{1}) \int dx \rho_{1}(x) \ln \rho_{1}(x) +
 (1 - {\lambda}_{2}) \int dx \rho_{2}(x) \ln \rho_{2}(x)
$$
$$
 - {\lambda}_{1} \int dx dy \rho_{1}(x) \ln |x - y | \rho_{1}(y) -
 {\lambda}_{2} \int dx dy  \rho_{2}(x) \ln |x - y | \rho_{2}(y)
$$
\begin{equation}
  - 2 {\lambda}_{12} \int dx dy \rho_{1}(x) \ln |x - y |  \rho_{2}(y).
\end{equation}
It is more convenient to work with an Hamiltonian, which unlike (\ref{htilde}),
is hermitian with respect to the flat functional Cartesian measure
$\prod_x d\rho_1(x)\,d\rho_2(x)\,.$ This we achieve by means of the 
similarity transformation $\psi \rightarrow J^{\frac{1}{2}} \psi\,, 
\tilde H \rightarrow H_{coll} = J^{\frac{1}{2}} \tilde H J^{- \frac{1}{2}}\,,$
where the continuum {\em collective} Hamiltonian is 
$$ H_{coll} ~~~~~~~~~~~~~~~~~~~~~~~~~~~~~~~~~~~~~~~~~~~~~~~~~~~~~~~~~~~~~~~~~~~~~~~~~~~~~~~~~~~~~~~~~~~~~~~~~~~~~~~~~~~~~~~~~~~~~~~~~~~~~~~~~~~~~~~~~~~~~~~$$
$$ = \frac{1}{2 m_{1}} \int dx\, \pi_1'(x)\, \rho_{1}(x)\, \pi_1'(x) +
  \frac{1}{2 m_{1}} \int dx \rho_{1}(x) {\left( \frac{\lambda_{1} - 1}{2} 
\frac{\partial_{x} \rho_{1}}{\rho_{1}} +
\lambda_{1} \pv \int \frac{ dy \rho_{1}(y)}{x - y}  + \lambda_{12} \pv \int 
\frac{dy \rho_{2}(y)}{x - y} \right)}^{2}
$$
$$
 + \frac{1}{2 m_{2}} \int dx \,\pi_2'(x)\,\rho_{2}(x) \,\pi_2'(x) 
 + \frac{1}{2 m_{2}} \int dx  \rho_{2}(x) {\left( \frac{\lambda_{2} - 1}{2} 
\frac{\partial_{x} \rho_{2}}{\rho_{2}}
  + \lambda_{2} \pv \int \frac{ dy  \rho_{2}(y)}{x - y}
  + \lambda_{12} \pv \int \frac{dy \rho_{1}(y)}{x - y} \right)}^{2} 
$$
\begin{equation} \label{colham}
 +  H_{sing},~~~~~~~~~~~~~~~~~~~~~~~~~~~~~~~~~~~~~~~~~~~~~~~~~~~~~~~~~~~~~~~~~~~~~~~~~~~~~~~~~~~~~~~~~~~~~~~~~~~~~~~~~~~~~~~~~~~~~~~~~~~~~~~~~~~~~~~~~~~~~~~
\end{equation}
where $ \; H_{sing} \; $ denotes a singular boundary contribution\footnote{
Note that the singular coefficients of $\rho_1$ and $\rho_2$ in (\ref{hsing}) 
are in fact independent of $x$, as evidently, $\partial_{x} \left. \frac{P}
{x - y} \right|_{y = x} =  \partial_{x} \left. \frac{P}{x} \right|_{x = 0} = 
\epsilon^{-2}\,,$ where in the last step we used the standard representation 
$\frac{P}{x} = {x\over x^2 + \epsilon^2}\,.$} : 
$$
 H_{sing} =  -  \int dx \left(\frac{\lambda_{1}}{2 m_{1}}  \rho_{1}(x) +
               \frac{\lambda_{2}}{2 m_{2}} \rho_{2}(x)
 \right) \partial_{x} \left. \frac{P}{x - y} \right|_{y = x}  
$$
\begin{equation}\label{hsing}
-  \int dx \left(\frac{\lambda_{1} - 1}{4 m_{1}} +  \frac{\lambda_{2}
 - 1}{4 m_{2}} \right)
 {\partial_{x}}^{2} \left. \delta(x - y) \right|_{y = x},
\end{equation}
and $ \; P \;$ is the principal part symbol.

It is well-known \cite{Jevicki:1979mb} that to leading order in the ${1\over N_{1,2}}$ 
expansion, collective dynamics of our system (properly represented by 
well-behaved observables) is determined by the classical equations of motion 
resulting from (\ref{colham}). 

As a concluding remark, note that the ${\pax\rho_{1,2}\over\rho_{1,2}}$ terms
in (\ref{colham}) are independent of the normalization (\ref{conservation}) of 
$\rho_1$ and $\rho_2$. Thus, they are clearly of next-to-leading  order in the 
${1\over N_{1,2}}$ expansion relative to the other terms. Nevertheless, 
these terms play an important role in in dynamics of the model, e.g., in 
governing small fluctuations around the ground state (see Sections 4 and 5), 
in the soliton sector of the model \cite{Bardek:2005yx}, and in establishing the duality 
symmetries of (\ref{colham}), which we do in the next section.


\section{Duality transformations and symmetries}
In this section we introduce and study the duality transformation and the
corresponding symmetries of the collective-field Hamiltonian (\ref{colham}).
It is straightforward to check that the Hamiltonian (\ref{colham}) is invariant
under the following set of transformations\footnote{Note that (\ref{duality1})
and (\ref{duality2}) do not constitute a symmetry of the original Hamiltonian 
(\ref{h1}).} of the parameters:
\begin{equation} \label{duality1}
 \tilde{\lambda}_{1} = \frac{1}{\lambda_{1}}, \;\;\;\;  \tilde{\lambda}_{2} = 
\frac{1}{\lambda_{2}}, \;\;\;\;
\tilde{m}_{1} = - \frac{m_{1}}{\lambda_{1}}, \;\;\;\;
 \tilde{m}_{2} = - \frac{m_{2}}{\lambda_{2}}, \;\;\;\; \tilde{\lambda}_{12} = 
\frac{1}{\lambda_{12}}
\end{equation}
and of the operators:
\begin{equation}  \label{duality2}
 \tilde{\rho}_{1} = - \lambda_{1} \rho_{1}, \;\;\;\;  \tilde{\rho}_{2} = - 
\lambda_{2} \rho_{2}, \;\;\;\;
 \tilde{\pi}_{1} = - \frac{\pi_{1}}{\lambda_{1}}, \;\;\;\;  \tilde{\pi}_{2} = 
- \frac{\pi_{2}}{\lambda_{2}}\,.
\end{equation}
Derivation of these transformations, as well as of the other sets of 
transformations discussed below in this section is straightforward: Consider 
two two-family Calogero systems with collective Hamiltonians of the form 
(\ref{colham}). Assume that the tilded quantities, corresponding say, to the 
second system, are simple homogeneous functions of the untilded ones 
(allowing sign flips). Assume further that the tilded and untilded quantities 
appearing in (\ref{threebody}) have the {\em same} common value. The 
transformations (\ref{duality1}) and (\ref{duality2}) then follow as the 
unique solution of equating the two Hamiltonians, which is neither an 
identity, nor a trivial permutation. (The $(\lambda_i -1)\pax\rho_i/\rho_i$ 
terms in (\ref{colham}), are crucial in obtaining these transformations 
uniquely.)

Let us denote the set of transformations (\ref{duality1})
and (\ref{duality2}) by $T_{12}$. These transformations are canonical, as 
they preserve the commutation relations (\ref{canonical}). 
For obvious reasons, we refer to the transformations $T_{12}$ 
as the strong-weak coupling duality transformation. 
Thus, we see that our Hamiltonian, expressed in terms of the new tilded 
parameters and operators, is identical in form to the original one, but
with $ \; \lambda_{1} \; $ and $ \; \lambda_{2} \; $ and the 
inter-family coupling  $ \; \lambda_{12} \; $ turned into their 
reciprocal values; with  $ \; N_{1} \; $ and $ \; N_{2} \; $ turned, 
respectively,  into $ \; \tilde N_1 = - \lambda_{1} N_{1} \; $ and $ \tilde N_2 = 
\; - \lambda_{2} N_{2} \; $ and, finally, with masses $ \; m_{1} \; $ and 
$ \; m_{2} \; $ turned into $ \; - \frac{m_{1}}{\lambda_{1}}  \; $ 
and $ \; - \frac{m_{2}}{ \lambda_{2}} $.
The minus signs which occur in these identifications are all important:
By drawing analogy to a similar situation in the one-family case 
\cite{Minahan:1994ce, halzirn}, we interpret all negative values of
the parameters and densities as those pertaining to holes, or 
anti-particles. Now, strictly speaking, since $N_i$ and $\tilde N_i$ are integers, 
this interpretation is consistent only for rational couplings, as was 
discussed in \cite{Minahan:1994ce, halzirn}.

We further note that (\ref{colham}) is invariant also under 
two more sets of canonical duality transformations. The first one, which 
we denote by $T_1$, is comprised of the set of transformations of parameters
\begin{equation} \label{duality6}
 \tilde{\lambda}_{1} = \lambda_{1}, \;\;\;\;  \tilde{\lambda}_{2} = 
\frac{1}{\lambda_{2}}, \;\;\;\;
\tilde{m}_{1} = m_{1}, \;\;\;\;
 \tilde{m}_{2} = - \frac{m_{2}}{\lambda_{2}}, \;\;\;\;
 \tilde{\lambda}_{12} = - \frac{\lambda_{12}}{\lambda_{2}}
\end{equation}
and of the operators
\begin{equation} \label{duality7}
 \tilde{\rho}_{1} = \rho_{1}, \;\;\;\;  \tilde{\rho}_{2} = - \lambda_{2} 
\rho_{2}, \;\;\;\;
 \tilde{\pi}_{1} = \pi_{1}, \;\;\;\;  \tilde{\pi}_{2} = - 
\frac{\pi_{2}}{\lambda_{2}}\,.
\end{equation}
Negative values of masses, densities and momenta, as in the previous case,
refer to holes. These transformations map the two-family Calogero model of 
particles (positive $ \; m_{1}, m_{2}, \rho_{1} \; $ and $ \; \rho_{2} $ ) 
with inter-family interaction strength $\; \lambda_{12} \; $ into the
dual two-family Calogero model of particles $ \;( m_{1}, \rho_{1}) \;$
and holes $ \;( {\tilde{m}}_{2}, {\tilde{\rho}}_{2}) \;$ with the inter-family 
interaction strength $\; - \frac{\lambda_{12}}{\lambda_{2}} $.
The second (and last) set of duality symmetries of (\ref{colham}), which we 
denote by $T_2$, is obtained from (\ref{duality6}) and (\ref{duality7}) 
simply by permuting the family indices $ \; 1 \leftrightarrow 2\,. $

It is easy to check that the duality transformations $T_1$, $T_2$, $T_{12}$, 
together with the identity transformation $ I $, form an Abelian group 
under composition, in which each element squares to $ I $, and where
$T_1 T_2 = T_{12}$, $T_1 T_{12} = T_2$ and $T_2 T_{12} = T_1$. This is 
readily identified as Klein's four-group. The latter is isomorphic to 
$Z\!\!\! Z_2 \otimes Z\!\!\! Z_2\,,$ where the two 
$Z\!\!\! Z_2 $ factors are $\{ I, T_1\}$ and 
$\{ I, T_2\}\,.$ As we shall see in Section 7, this group-theoretic 
pattern of dualities persists also in the generic case.

\subsection{Resemblence of special two-family Calogero models to
single-family Calogero models}
As an interesting application of the duality symmetry group, consider the 
special case of the two-family Calogero model (\ref{h1}) in 
which\footnote{If we ignore inter-family coupling, we can think of this 
system as made of two single-family models, related by the one-family 
version of the strong-weak coupling duality, save for the relation between 
densities.} $\lambda_2 = {1\over \lambda_1}$ and 
$m_2 = - {m_1\over \lambda_1}\,.$
From (\ref{threebody}) we then find that $\lambda_{12} = {m_2\over m_1}
\lambda_1 = -1$. Note that the two particle families in this system are 
generically manifestly distinct. Nevertheless, this distinction is, in some 
sense, an illusion. To see this, note that by the duality transformations 
(\ref{duality6}) and (\ref{duality7}), namely, the element $T_1$ of the 
duality group, this system is equivalent to a two-family system with 
parameters $\tilde{\lambda}_{1} = \tilde{\lambda}_{2} =  \tilde{\lambda}_{12} 
= \lambda_{1}\,,\quad \tilde{m}_{1} =  \tilde{m}_{2} = m_1 $
and densities $ \tilde{\rho}_{1} = \rho_{1}\,,\tilde{\rho}_{2} = - 
{\rho_{2}\over \lambda_1}\,.$ In the latter dual system, the two families are 
identical! For this reason, we may refer to the two dimensional locus 
\beq\label{democratic}
\lambda_2 = {1\over \lambda_1}\,,\quad\quad m_2 = - {m_1\over \lambda_1}\,,
\quad\quad \lambda_{12} = -1
\eeq
in parameter space as the ``surface of hidden identity'', or SOHI.
Thus, the special two-family Calogero model we 
started with resembles the single-family Calogero model specified by  
\begin{equation} \label{ftn1}
\lambda = {\lambda}_{1}\,, \quad\quad m = m_{1}\,, \quad\quad
\rho = \tilde\rho_1 + \tilde\rho_2 = \rho_1 - \frac{1}{\lambda_1}{\rho_2}\,.
\end{equation}
Similarly, by inverting the roles of family indices $1\leftrightarrow 2$
in the previous discussion, which leaves us on the SOHI  
(\ref{democratic}), and then applying the duality tranformation $T_2$, 
we shall conclude that the special two-family Calogero model we started with 
resembles the single-family Calogero model specified by  
\begin{equation} \label{ftn2}
\lambda = {\lambda}_{2}\,, \quad\quad m = m_{2}\,, \quad\quad
\rho = \rho_2 - \frac{1}{\lambda_2}{\rho_1}\,.
\end{equation}
In both cases, the effective single-species collective field $ \; \rho
\; $ actually shares the statistics $ \; \lambda \; $ and the mass $\; m \; $
with the first or the second family, respectively. Note that these two cases
can be mapped one each other by the duality transformation $T_{12}$. Thus, 
the SOHI (\ref{democratic}) is left invariant under $T_{12}$. 
However, the latter does not act on it freely, as $\lambda_1=\lambda_2 = 
-\lambda_{12} = 1$ and $m_1 = -m_2$ is a fixed line. Models lying on this
line are comprised of particles and their antiparticles, and only particles 
and antiparticles interact (repulsively).

Note that we described the relation between the original special two-family 
models and the corresponding single-family models merely as ``resemblance''. 
They are certainly {\em not} identical! The density operator $\rho$ 
appearing on the LHS of (\ref{ftn1}) and (\ref{ftn2}), which corresponds to
the single-family Calogero model, is defined in a Hilbert space made
of many-body wave functions which are completely symmetric in the coordinates
of all particles. $\rho_1$ and $\rho_2$, on the other hand, are symmetric 
only in the coordinates of particles of each family separately.
The best one could do is perhaps to consider the two-family system
with identical families (the one dual to the special two-family 
systems we started with) as a one-family system divided into two parts, 
differing by some internal quantum number, in which one symmetrizes in each 
sector separately. However, this means one should also contrapt an 
actual physical context to justify such separate symmetrization.


\section{Ground state and small fluctuations}
In this section we shall concentrate on the ground-state of the two-family
Calogero system and the small excitations above it. To this end, 
it is sufficient to consider only static density fields, for which, 
of course, $\pi_1=\pi_2 = 0$. Evaluating the Hamiltonian 
(\ref{colham}) on such fields and imposing the normalization contraints 
(\ref{conservation}), yields the effective potential, given 
(to leading order in the $ \; 1/N_{1} \; $ and $ \; 1/N_{2} \; $ expansion) by 
$$ 
 V  = \frac{1}{2 m_{1}} \int dx \rho_{1}(x) {\left( \frac{\lambda_{1} - 1}{2} 
\frac{\partial_{x} \rho_{1}}{\rho_{1}} + 
 \lambda_{1} \pv \int \frac{ dy \rho_{1}(y)}{x - y}  + \lambda_{12} \pv \int 
\frac{dy \rho_{2}(y)}{x - y} \right)}^{2} 
$$
$$ 
+  \frac{1}{2 m_{2}} \int dx  \rho_{2}(x) {\left( \frac{\lambda_{2} - 1}{2} \frac{\partial_{x} \rho_{2}}{\rho_{2}}
  + \lambda_{2} \pv \int \frac{ dy \rho_{2}(y)}{x - y}
  + \lambda_{12} \pv \int \frac{dy \rho_{1}(y)}{x - y} \right)}^{2}
$$
\begin{equation}\label{potential1}
+ \mu_{1} \left(N_{1} - \int dx \rho_{1}(x) \right) + \mu_{2} \left(N_{2} - 
\int dx \rho_{2}(x) \right)\,.
\end{equation}
Here  $ \; \mu_{1} \; $ and $ \; \mu_{2} \; $ are the chemical potentials 
which impose (\ref{conservation}).\footnote{In (\ref{potential1}) we have 
absorbed the constant singular coefficients of $\rho_1$ and $\rho_2$ 
which appear in $H_{sing}$ in (\ref{hsing}) as infinite shifts
in $\mu_1$ and $\mu_2$, and also omitted the infinite, density independent 
constant in $H_{sing}$.}

The expression in (\ref{potential1}) contains bilocal as well as trilocal 
terms in the densities. We can avoid the menacing trilocal terms by applying 
standard tricks as follows: 
As was mentioned at the beginning of the previous section, we tacitly 
assume that the system is properly regularized at large distances, meaning
in particular, that $\rho_1(x)$ and $\rho_2(x)$ are always of compact support. 
The principal value distribution, acting on such functions, satisfies 
the identity 
\begin{equation}
\frac{P}{x-y} \frac{P}{x-z} + \frac{P}{y-z} \frac{P}{y-x} + 
\frac{P}{z-x} \frac{P}{z-y} =   {\pi}^{2} \delta(x-y) \delta(x-z)\,.
\label{principal}
\end{equation}
Making use of (\ref{principal}) in (\ref{potential1}), we obtain
$$
 V = \frac{{\pi}^{2}}{6 \lambda_{1} m_{1}} \int dx {(\lambda_{1}
 \rho_{1} + \lambda_{12} \rho_{2})}^{3} +
$$
$$
     + \frac{{(\lambda_{1} - 1)}^{2}}{8 m_{1}} \int dx \frac{{(\partial_{x} 
\rho_{1})}^{2}}{\rho_{1}} +
     \frac{{(\lambda_{2} - 1)}^{2}}{8 m_{2}} \int dx \frac{{(\partial_{x} 
\rho_{2})}^{2}}{\rho_{2}}
$$
$$
 + \frac{ \lambda_{1} (\lambda_{1} - 1)}{2 m_{1}} \int dx \; \partial_{x} 
\rho_{1} \;
             \pv \int dy \frac{\rho_{1}(y)}{x-y}  
            + \frac{ \lambda_{2} (\lambda_{2} - 1)}{2 m_{2}} \int dx \; 
\partial_{x} \rho_{2} \;
             \pv \int dy \frac{\rho_{2}(y)}{x-y} 
$$
$$
 + \frac{ \lambda_{12} (\lambda_{1} - 1)}{2 m_{1}} \int dx \; \partial_{x} 
\rho_{1} \;
             \pv \int dy \frac{\rho_{2}(y)}{x-y} 
            + \frac{ \lambda_{12} (\lambda_{2} - 1)}{2 m_{2}} \int dx \; 
\partial_{x} \rho_{2} \;
             \pv \int dy \frac{\rho_{1}(y)}{x-y}.
$$
\begin{equation}\label{vofh}
+ \mu_{1} \left(N_{1} - \int dx \rho_{1}(x) \right) + \mu_{2} \left(N_{2} - 
\int dx \rho_{2}(x) \right)\,.
\end{equation}
This expression for $V$ is evidently devoid of trilocal terms. In Appendix A 
we present an alternative derivation of (\ref{vofh}). 

As was mentioned at the end of the previous section, in the large-$N$ limit, 
dynamics of the system is semiclassical, governed 
by the saddle points of (\ref{colham}). In particular, 
the ground-state energy and the corresponding density distributions can be 
found, to leading order in ${1\over N_{1,2}}$, by varying the effective 
potential (\ref{vofh}) with respect to $\rho_{1}$ and $\rho_{2}, $ which  
yields the two coupled integro-differential, non-linear equations
$$
\frac{{\pi}^{2}}{2 m_{1}} {(\lambda_{1} \rho_{1} + \lambda_{12} 
\rho_{2})}^{2} -
\frac{{(\lambda_{1} - 1)}^{2}}{8 m_{1}} { ( \frac{\partial_{x} \rho_{1}}
{\rho_{1}})}^{2} -
\frac{{(\lambda_{1} - 1)}^{2}}{4 m_{1}} \partial_{x} ( \frac{\partial_{x} 
\rho_{1}}{\rho_{1}}) 
$$
\begin{equation} \label{br1}
- \frac{\lambda_{1} (\lambda_{1} - 1)}{m_{1}}  \pv \int dy \frac{\partial_{y} 
\rho_{1}(y)}{x-y} - \frac{\lambda_{12}}{2} ( \frac{\lambda_{1} - 1}{m_{1}} + 
\frac{\lambda_{2} - 1}{m_{2}})
   \pv \int dy \frac{\partial_{y} \rho_{2}(y)} {x-y} = \mu_{1},
\end{equation}
 $$
 \frac{{\pi}^{2}}{2 m_{2}} {(\lambda_{2} \rho_{2} + \lambda_{12} 
\rho_{1})}^{2} -
\frac{{(\lambda_{2} - 1)}^{2}}{8 m_{2}} { ( \frac{\partial_{x} \rho_{2}}
{\rho_{2}})}^{2} - \frac{{(\lambda_{2} - 1)}^{2}}{4 m_{2}} \partial_{x} 
( \frac{\partial_{x} \rho_{2}}{\rho_{2}}) 
$$
\begin{equation} \label{br2}
-  \frac{\lambda_{2} (\lambda_{2} - 1)}{m_{2}}  \pv \int dy 
\frac{\partial_{y} \rho_{2}(y)}{x-y} -
\frac{\lambda_{12}}{2} ( \frac{\lambda_{1} - 1}{m_{1}} + 
\frac{\lambda_{2} - 1}{m_{2}})
   \pv \int dy \frac{\partial_{y} \rho_{1}(y)} {x-y} = \mu_{2}.
\end{equation}
In the ground state, the two particle types evidently condense into uniform 
density configurations 
\begin{equation} \label{unifconfig}
 \rho_{1,0} = \frac{N_{1}}{L}\,,\quad\quad \rho_{2,0} = \frac{N_{2}}{L}\,,
\end{equation}
inside the spatial box of length $L$, which confines the particles. 
These uniform densities are clearly solutions of the variational equations 
(\ref{br1}) and (\ref{br2}), since for uniform densities the latter equations
reduce simply to relations which determine the chemical potentials. 
To conclude our discussion of the semiclassical limit, note that to leading 
order in ${1\over N_{1,2}}$, the ground-state energy corresponding 
to (\ref{unifconfig}) follows from (\ref{vofh}):
\begin{equation} \label{e1}
 E_{0} = \frac{{\pi}^{2}}{6 \lambda_{1} m_{1}} \int dx 
          {(\lambda_{1} \rho_{1,0} + \lambda_{12} \rho_{2,0})}^{3}
   =  \frac{{\pi}^{2}}{6 \lambda_{1} m_{1} {L}^{2}}  
          {(\lambda_{1} N_{1} + \lambda_{12} N_{2})}^{3}.
\end{equation}
This is, in fact, the exact result for finite $ N_{1}$ and $ N_{2} $ \cite{Sen:1995zj}.
It is worth mentioning here that (\ref{br1}) and (\ref{br2}) possess also 
nonuniform static soliton solutions, which were studied extensively 
in \cite{Bardek:2005yx, Bardek:2005cs}.

The collective-field formalism provides a systematic framework for the 
$ {1\over N_{1,2}}$ expansion \cite{Andric:1991wp, Andric:1994su}. In particular, by expanding around the
classical uniform configurations (\ref{unifconfig}) (with their corresponding
null momenta), we can go beyond the leading order and obtain the spectrum of 
low-lying excitations above the ground state.
Here we shall concentrate on the next-to-leading terms. To this end we write
\begin{equation} \label{fluctuations}
\rho_{1} (x) = \rho_{1,0} + \eta_{1} (x)\,,\quad\quad
\rho_{2} (x) = \rho_{2,0} + \eta_{2} (x)\,,
\end{equation}
where $\eta_{1} (x) $ and $ \eta_{2} (x) $ are small density fluctuations, with 
$\eta_a$ being typically of order $1/N_a$. 
Substituting (\ref{fluctuations}) in the Hamiltonian (\ref{colham}), and 
expanding it to quadratic order in
$\pi_{1}, \pi_{2}, \eta_{1}\,,$ and $\eta_{2}\,,$ we obtain 
$$
H_{coll} = \frac{\rho_{1,0}}{2 m_{1}} \int dx  { ( \partial_{x} 
\pi_{1}(x))}^{2} + \frac{\rho_{1,0}}{2 m_{1}} \int dx  
{\left( \frac{\lambda_{1} - 1}{2} \frac{\partial_{x} \eta_{1}}{\rho_{1,0}} +
\lambda_{1} \pv \int \frac{ dy \eta_{1}(y)}{x - y}  + \lambda_{12} \pv \int 
\frac{dy \eta_{2}(y)}{x - y} \right)}^{2}
$$
$$
 + \frac{\rho_{2,0}}{2 m_{2}} \int dx {(\partial_{x} \pi_{2}(x))}^{2} 
 + \frac{\rho_{2,0}}{2 m_{2}} \int dx  {\left( \frac{\lambda_{2} - 1}{2} 
\frac{\partial_{x} \eta_{2}}{\rho_{2,0}}
  + \lambda_{2} \pv \int \frac{ dy  \eta_{2}(y)}{x - y}
  + \lambda_{12} \pv \int \frac{dy \eta_{1}(y)}{x - y} \right)}^{2}
$$
\begin{equation} \label{colham1}
 + H_{sing} + E_{0}\,.
\end{equation}
The quadratic approximation (\ref{colham1}) to $H_{coll}$, as the original 
collective Hamiltonian (\ref{colham}), is invariant under the duality 
transformations $T_1$, $T_2$ and $T_{12}$ in Section 3.
Eq. (\ref{colham1}) governs the leading-order fluctuations of the system 
around the ground state. Its quadratic structure clearly calls for 
introducing the  two operators
\begin{equation} \label{grrr1}
 A_{1}(x) =  \partial_{x} \pi_{1}(x) + i \left( \frac{\lambda_{1} - 1}{2} 
\frac{\partial_{x} \eta_{1}}{\rho_{1,0}} +
 \lambda_{1} \pv \int \frac{ dy \eta_{1}(y)}{x - y}  + \lambda_{12} \pv \int 
\frac{dy \eta_{2}(y)}{x - y} \right),
\end{equation}
\begin{equation} \label{grrr2}
 A_{2} (x) = \partial_{x} \pi_{2}(x) + i \left( \frac{\lambda_{2} - 1}{2} 
\frac{\partial_{x} \eta_{2}}{\rho_{2,0}}
  + \lambda_{2} \pv \int \frac{ dy  \eta_{2}(y)}{x - y}
  + \lambda_{12} \pv \int \frac{dy \eta_{1}(y)}{x - y} \right),
\end{equation}
along with their hermitian adjoints. Evidently, these operators are the
key to computing the collective quasiparticle spectrum of small fluctuations
around the ground state. Higher order terms in the expansion of $H_{coll}$ 
around the uniform densities yield the interactions among 
these quasiparticles.

It can be checked that the duality transformations
of Section 3 act on $A_1(x)$ and $A_2(x)$ as follows: 
$A_1$ is invariant under $T_1$ while $T_1 A_2 \rightarrow -{1\over\lambda_2} 
A_2\,;$ $A_2$ is invariant under $T_2$ while 
$T_2 A_1 \rightarrow -{1\over\lambda_1} A_1\,;$ finally, 
$T_{12} A_{1,2} \rightarrow -{1\over\lambda_{1,2}} A_{1,2}\,$ (the latter
transformation is consistent with the group multiplication 
$T_1 T_2 = T_{12}$\,).

The operators (\ref{grrr1}), (\ref{grrr2}) and their adjoints satisfy the 
commutation relations
$$
 [A_{1}(x), A_{1}^{\dagger}(y)] = \frac{1 - \lambda_{1}}{\rho_{1,0}} 
\partial_{x}\partial_{y} \delta(x - y)
                                  + 2\lambda_{1} \partial_{x} \frac{P}{x - y},
$$
\begin{equation}\label{Acommutators}
 [A_{2}(x), A_{2}^{\dagger}(y)] = \frac{1 - \lambda_{2}}{\rho_{2,0}} 
\partial_{x}\partial_{y} \delta(x - y)
                                  + 2\lambda_{2} \partial_{x} \frac{P}{x - y},
\end{equation}
$$
[A_{2}(x), A_{1}^{\dagger}(y)] =  [A_{1}(x), A_{2}^{\dagger}(y)] = 
2 \lambda_{12} \partial_{x} \frac{P}{x - y},
$$
with all other commutators vanishing. In terms of these operators the 
quadratic Hamiltonian (\ref{colham1}) takes the form
$$
H_{coll} = E_{0} + \frac{\rho_{1,0}}{2 m_{1}} \int dx A_{1}^{\dagger}(x)
A_{1}(x) + 
     \frac{\rho_{2,0}}{2 m_{2}} \int dx A_{2}^{\dagger}(x)A_{2}(x) 
$$
$$
    + \frac{\rho_{1,0}}{4 m_{1}} \int dx [A_{1}(x), A_{1}^{\dagger}(x)] +
     \frac{\rho_{2,0}}{4 m_{2}} \int dx [A_{2}(x), A_{2}^{\dagger}(x)] + 
H_{sing}\,. 
$$
As can be seen from (\ref{Acommutators}), the commutator terms in the 
last equation, which are essentially the sum over zero-point energies 
originating from the quadratic pieces in (\ref{colham1}), cancel the 
the singular term $H_{sing}$ defined in (\ref{hsing}). This is a typical 
behavior in collective field theory. Thus, we obtain the finite result
\begin{equation} \label{colham2}
H_{coll} = E_{0} + \frac{\rho_{1,0}}{2 m_{1}} \int dx A_{1}^{\dagger}(x)
A_{1}(x) + 
     \frac{\rho_{2,0}}{2 m_{2}} \int dx A_{2}^{\dagger}(x)A_{2}(x).
\end{equation}
As our next step towards diagonalization of (\ref{colham2}), we move to 
momentum space and express the operators $A_{1}$ and $ A_{2} $ in terms of
of their Fourier transforms
\beq\label{Fourier}
A_{1}(x) = \sqrt{\frac{m_{1}}{\pi \rho_{1,0}}} 
\int\limits_{-\infty}^\infty dk e^{i k x} a_{1}(k)\,,
\quad\quad 
A_{2} (x) = \sqrt{\frac{m_{2}}{\pi \rho_{2,0}}} \int\limits_{-\infty}^\infty
 dk e^{i k x} a_{2}(k)\,.
\end{equation}
By definition of the duality trasnformations in Section 3, and from their 
action on $A_1$ and $A_2$ as described following (\ref{grrr2}), we obtain 
that their action on $a_1$ and $a_2$ reduces to either flipping a sign or not:
$a_1$ is invariant under $T_1$ while $T_1 a_2 \rightarrow -a_2\,;$ $a_2$ is 
invariant under $T_2$ while $T_2 a_1 \rightarrow -a_1\,;$ finally, 
$T_{12} a_{1,2} \rightarrow -a_{1,2}\,.$

>From (\ref{grrr1}) and (\ref{grrr2}), and from the identity 
\begin{equation}
\label{pv} 
\pv\int_{-\infty}^\infty dx \frac{ e^{ikx}}{x-y} = i \pi e^{iky}\,{\rm sign}
\, k 
\end{equation}
it is easy to verify that 
\beqra\label{aofk}
a_1(k) &=& \sqrt{{\pi\rho_{1,0}\over m_1}}\,\left[ ik\tilde\pi_1(k) + 
{1-\lambda_1\over 2\rho_{1,0}} k \tilde\eta_1 (k) + \pi\left(\lambda_1
\tilde\eta_1(k) + \lambda_{12}\tilde\eta_2(k)\right){\rm sign}\,k\right]
\nonumber
\\{}\nonumber\\
a_2(k) &=& \sqrt{{\pi\rho_{2,0}\over m_2}}\,\left[ ik\tilde\pi_2(k) + 
{1-\lambda_2\over 2\rho_{2,0}} k \tilde\eta_2 (k) + \pi\left(\lambda_2
\tilde\eta_2(k) + \lambda_{12}\tilde\eta_1(k)\right){\rm sign}\,k\right]\,.
\nonumber\\{}
\eeqra
The Fourier modes  
\beq\label{Fourier1}
\tilde\eta_a (k) = \int\limits_{-\infty}^\infty {dx\over 2\pi} 
e^{-i k x} \eta_{a}(x)\,,
\quad\quad 
\tilde \pi_{a} (k) = \int\limits_{-\infty}^\infty {dx\over 2\pi} 
e^{-i k x} \pi_{a}(x)
\end{equation}
obey the commutation relations\footnote{Due to the fact that 
$\int\limits_{-\infty}^\infty\,dx \,\eta_a(x) = 0$, the zero-momentum modes
of the fluctuation fields are nondynamical commuting numbers. See the 
footnote just above (\ref{canonical}) for a similar matter concerning the 
full density fields. As was commented upon there, in the limit of very large 
spatial box, we can safely use (\ref{canonical1}).}'\footnote{Note that 
$\tilde\eta_a^\dgg (k) = \tilde\eta_a (-k)$ and $\tilde\pi_a^\dgg (k) = 
\tilde\pi_a (-k)$, since $\eta_a(x)$ and $\pi_a(x)$ are hermitian field 
operators.}
\beq\label{canonical1}
[\tilde\eta_a (k), \tilde\pi_b (q)] = {i\over 2\pi}\,\delta (k+q)\,,
\eeq
with all other commutators being null. For convenience, we also write down 
the hermitian adjoints of (\ref{aofk}):
\beqra\label{adggofk}
a_1^\dgg (k) &=& \sqrt{{\pi\rho_{1,0}\over m_1}}\,\left[ -ik\tilde\pi_1(-k) + 
{1-\lambda_1\over 2\rho_{1,0}} k \tilde\eta_1 (-k) + \pi\left(\lambda_1
\tilde\eta_1(-k) + \lambda_{12}\tilde\eta_2(-k)\right){\rm sign}\,k\right]
\nonumber
\\{}\nonumber\\
a_2^\dgg (k) &=& \sqrt{{\pi\rho_{2,0}\over m_2}}\,\left[ -ik\tilde\pi_2(-k) + 
{1-\lambda_2\over 2\rho_{2,0}} k \tilde\eta_2 (-k) + \pi\left(\lambda_2
\tilde\eta_2(-k) + \lambda_{12}\tilde\eta_1(-k)\right){\rm sign}\,k\right]\,.
\nonumber\\{}
\eeqra
Note that by combining the equations for $a_b(k)$ and $a_b^\dgg(-k)$ we can 
invert (\ref{aofk}) and (\ref{adggofk}), and express $\eta_b(k)$ and 
$\pi_b(k)$ as linear combinations of $a_b(k)\pm a_b^\dgg (-k)\,.$

Finally, it follows from (\ref{aofk}), (\ref{canonical1}) and 
(\ref{adggofk}) that\footnote{Alternatively, Eqs. (\ref{commutator1}) - 
(\ref{commutator}) can be verified by Fourier-transforming (\ref{Acommutators})
directly.}
\begin{equation} \label{commutator1}
 [a_{1}(k), a_{1}^{\dagger}(k')] = \left( \frac{1 - \lambda_{1}}{2 m_{1}} k^{2}
                           + \frac{ \lambda_{1} \pi \rho_{1,0}}{m_{1}}|k| 
\right) \delta(k - k')
    \equiv \omega_{1}(k) \delta(k - k'),
\end{equation}
\begin{equation} \label{commutator4}
 [a_{2}(k), a_{2}^{\dagger}(k')] = \left( \frac{1 - \lambda_{2}}{2 m_{2}} k^{2}
                           + \frac{ \lambda_{2} \pi \rho_{2,0}}{m_{2}}|k| 
\right) \delta(k - k')
    \equiv \omega_{2}(k) \delta(k - k'),
\end{equation}
\begin{equation} \label{commutator}
 [a_{1}(k), a_{2}^{\dagger}(k')] = [a_{2}(k), a_{1}^{\dagger}(k')] =  
\lambda_{12} \pi \sqrt{\frac{\rho_{1,0} \rho_{2,0}}{m_{1} m_{2}}} |k|  
\delta(k - k') \equiv \omega_{12}(k) \delta(k - k'),
\end{equation}
and all other commutators vanish. For later use, it will be 
convenient to lump (\ref{commutator1}) - (\ref{commutator}) into a real 
symmetric matrix $[a_a(k), a_b^\dgg (k')] = \omega_{ab} (k) \delta(k - k')\,,$
with 
\begin{equation} \label{omega}
 {\bf \omega} \equiv
\left(
\begin{array}{cc}
  \omega_{1}(k)   &   \omega_{12}(k)  \\
  \omega_{12}(k)  &   \omega_{2}(k)
\end{array}
\right).
\end{equation}
It is easy to verify that the duality transformations of Section 3 act on 
${\bf \om}$ as follows: $T_{12}$ leaves it invariant, while both $T_1$ and 
$T_2$ transform it to $\sigma_3 {\bf \om} \sigma_3$ (i.e., they flip the 
sign of the off-diagonal terms $\om_{12}$). All this is of course consistent
with the action of these duality transformations on $a_1$ and $a_2$, as 
described following (\ref{Fourier}). Note in passing that we can infer a 
special case of the exact duality equivalence of systems with parameters on 
the SOHI (\ref{democratic}) and two-family models made of identical families, 
as discussed in Section 3.1. This we achieve simply by demanding that 
$\om_{11}(k) = \om_{22} (k)$, which immediately leads to the pair of 
equations $\frac{1 - \lambda_{1}}{m_{1}} = 
\frac{1 - \lambda_{2}}{m_{2}}$ and $\frac{ \lambda_{1}\rho_{1,0}}{m_{1}}
= \frac{ \lambda_{2} \rho_{2,0}}{m_{2}}\,.$ Discarding the trivial solution 
(i.e., that the two families are identical to begin with), we obtain that the
conditions defining the SOHI (\ref{democratic}) comprise a solution of the
first equation, which upon substitution into the second equation leads
to the further relation $ \rho_{2,0} = - \lambda_1  \rho_{1,0}$\,
(which did not arise in Section 3.1).

In terms of the operators (\ref{aofk}) and (\ref{adggofk}) the Hamiltonian 
(\ref{colham2}) becomes
\begin{equation} \label{colham3}
 H_{coll} = E_{0} + \int dk  a_{1}^{\dagger}(k)a_{1}(k) +
   \int dk   a_{2}^{\dagger}(k)a_{2}(k).
\end{equation}
In Section 5 we will show that the ground state of the fluctuation fields 
is annihiliated by $a_1(k)$ and $a_2(k)$. Thus, the ground-state energy 
$ \; E_{0} \; $ is not affected by the next-to-leading-order corrections 
stemming from the quantum collective-field fluctuations 
$ \; \eta_{1} \; $ and $ \; \eta_{2}. $


\section{Diagonalization and dispersion laws}

Expression (\ref{colham3}) is not really the diagonal form of the 
Hamiltonian, since the operators $a_{1}$ and $a_{2}$ do not commute - see 
(\ref{commutator}). As usual, complete decomposition of (\ref{colham3}) into 
independent normal modes can be achieved by properly rotating  $a_{1} (k)$ and 
$a_{2} (k)$ into  two new {\em commuting} bosonic operators $b_{1} (k)$ and 
$b_{2} (k)$. Thus, we write 
\begin{equation} \label{orttr}
 b_a (k) = \sum_{b = 1}^{2} u_{ab}(k)\, a_b(k)\,,
\end{equation}
where $u_{ab}(k)$ are c-number coefficients. They are chosen such that 
the matrix ${\bf u}$ diagonalizes the commutator matrix (\ref{omega}). Since 
${\bf \omega}$ is real and symmetric, ${\bf u}$ is real orthogonal. Thus, 
\begin{equation}\label{diagonalization}
  {\bf  u} \; {\bf \omega} \; {{\bf u}^T} = {\bf \Omega}\,,
\end{equation}
where 
\begin{equation}\label{Omega}
 {\bf \Omega} \equiv
\left(
\begin{array}{cc}
  \Omega_{1}(k)   &   0  \\
       0          &   \Omega_{2}(k)
\end{array}
\right).
\end{equation}
Recall that the duality transformations $T_1$, $T_2$ and $T_{12}$ transform
${\bf\om}$ either to itself or to $\sigma_3 {\bf\om}\sigma_3\,,$ which is 
similar to ${\bf\om}$\,. Thus, the the eigenvalues $\Omega_a (k)$ of 
(\ref{omega}) are {\em invariant} under these duality transformations. These
eigenvalues are given by the standard 
formula
\begin{equation} \label{dispersion}
 \Omega_{1,2}(k) = \frac{1}{2} \left( \omega_{1}(k) + \omega_{2}(k) \pm
               \sqrt{{(\omega_{1}(k) - \omega_{2}(k))}^{2} + 4 {\omega_{12}}^{2}(k)} \right).
\end{equation}
By construction, therefore, the new operators satisfy the decoupled 
commutation algebra
$$
 [b_{1}(k), b_{1}^{\dagger}(k')] = \Omega_{1}(k) \delta(k - k'),
$$
\begin{equation} \label{grm}
 [b_{2}(k), b_{2}^{\dagger}(k')] = \Omega_{2}(k) \delta(k - k'),
\end{equation}
$$
 [b_{1}(k), b_{2}^{\dagger}(k')] =  0.
$$
The coefficients $u_{ab}(k)$ of the orthogonal transformation can be 
expressed in terms of the rotation angle $\phi (k)$ as 
\beqra\label{phik}
u_{11} = u_{22} = \cos\,\phi (k) &=& {2\om_{12}\over \sqrt{4\om_{12}^2 + 
[\om_2-\om_1 + \sqrt{4\om_{12}^2 + (\om_2-\om_1)^2}\,]^2}}\nonumber\\
{}\nonumber\\
u_{12} = -u_{21} = \sin\,\phi (k) &=& {\om_2-\om_1 + 
\sqrt{4\om_{12}^2 + (\om_2-\om_1)^2}}\over \sqrt{4\om_{12}^2 + 
[\om_2-\om_1 + \sqrt{4\om_{12}^2 + (\om_2-\om_1)^2}\,]^2}\,.
\eeqra
Note from (\ref{commutator1}) - (\ref{commutator}) that the $u_{ab}(k)$ 
are all even functions of $k$, in addition to being real. 
A useful and simple identity which we mention in passing is 
\begin{equation}\label{rotation}
 \cot 2 \phi = \frac{\omega_{1} - \omega_{2}}{2 \omega_{12}}.
\end{equation}
By rescaling
\begin{equation}
 b_{1}(k) \rightarrow \sqrt{\Omega_{1}(k)} \;\; b_{1}(k)\,,\quad\quad 
 b_{2}(k) \rightarrow \sqrt{\Omega_{2}(k)} \;\; b_{2}(k),
\end{equation} 
we finally obtain the diagonal Hamiltonian
\begin{equation} \label{grrr}
 H_{coll} = E_{0} + \int dk \Omega_{1}(k) b_{1}^{\dagger}(k)b_{1}(k) +
   \int dk \Omega_{2}(k)  b_{2}^{\dagger}(k)b_{2}(k)\,.
\end{equation}
$\Omega_{1}(k)$ and $\Omega_{2}(k)$ are therefore the energy spectra
of the physical fluctuations, i.e. the two decoupled collective oscillator 
modes, or quasiparticles of the system.

As can be readily seen from the commutation algebra
 (\ref{commutator1})-(\ref{commutator}), the dispersions $\omega_{1}(k)$ and 
$\omega_{2}(k)$, evaluated at $\lambda_{12} = 0\,,$ actually represent the
low-energy dispersions for the two non-interacting 
one-family Calogero systems. Stability of these latter two systems requires 
that the coefficients of the terms in  $\omega_{1}(k)$ and 
$\omega_{2}(k)$ which are proportional to $|k|$, the leading term in the 
low-momentum behavior of these systems, be positive. Thus, we shall henceforth 
make the assumption that these terms are positive. Note further that these terms 
are invariant under the duality transformations of Section 3, and so are their signs.

Since we know from \cite{Sen:1997qt}
that the low-energy one-family dispersions $\omega_{1}(k)$ and $\omega_{2}(k)$
are correctly given only up to $ k^{2} $ (near $ k = 0) $, the same must be 
also true for $\Omega_{1}(k) $ and $\Omega_{2}(k)$. By expanding
the square root in (\ref{dispersion}) up to order $k^2$, we 
obtain\footnote{As was stated above, here we assume that the coefficient 
of $|k|$ in the first equation in (\ref{dispersion1}) is positive. In case it 
is negative, the term linear in $|k|$ would appear in $\Om_2(k)$ instead of 
$\Om_1(k)$. }
\beqra\label{dispersion1}
\Omega_{1}(k) &=& 
\frac{ (1 - \lambda_{1})\lambda_{1} \rho_{1,0} /{m_{1}}^{2} +
(1 - \lambda_{2})\lambda_{2} \rho_{2,0} /{m_{2}}^{2}}
{  \lambda_{1} \rho_{1,0} / m_{1} +   \lambda_{2} \rho_{2,0} / m_{2}} 
\frac{k^{2}}{2} +  \pi \left( \frac{ \lambda_{1} \rho_{1,0}}{m_{1}} +  
\frac{ \lambda_{2} \rho_{2,0}}{m_{2}} \right) |k|\,,\nonumber\\{}\nonumber\\
\Omega_{2}(k) &=& 
\frac{ (1 - \lambda_{2})\lambda_{1} \rho_{1,0} / (m_{1} m_{2}) +
(1 - \lambda_{1})\lambda_{2} \rho_{2,0} / (m_{1} m_{2}) }
{  \lambda_{1} \rho_{1,0} / m_{1} +   \lambda_{2} \rho_{2,0} / m_{2}}   
\frac{k^{2}}{2 }\,.
\eeqra
The first dispersion $ \; \Omega_{1}(k) \; $ has the same structure as
 that of a single-species Calogero model \cite{Andric:1994su} with 
some effective particle density $ \; \rho, \; $ mass  $ \; m \; $ and
interaction parameter $ \; \lambda \; $ satisfying 
the following two relations:
\begin{equation} \label{con1}
 \frac{ (1 - \lambda_{1})\lambda_{1} \rho_{1,0} /{m_{1}}^{2} +
      (1 - \lambda_{2})\lambda_{2} \rho_{2,0} /{m_{2}}^{2}}
    {  \lambda_{1} \rho_{1,0} / m_{1} +   \lambda_{2} \rho_{2,0} / m_{2}} = 
\frac{1 - \lambda}{ m},
\end{equation}
\begin{equation} \label{con2}
 \frac{ \lambda_{1} \rho_{1,0}}{m_{1}} +  
\frac{ \lambda_{2} \rho_{2,0}}{m_{2}} = \frac{\lambda \rho}{m}.
\end{equation}
The second dispersion $ \; \Omega_{2}(k) \; $ describes a free-particle. 
The effective mass $ \; m^{*}  \; $ of this quasi-particle is given by
\begin{equation}\label{mstar}
 \frac{ (1 - \lambda_{2})\lambda_{1} \rho_{1,0} / (m_{1} m_{2}) +
      (1 - \lambda_{1})\lambda_{2} \rho_{2,0} / (m_{1} m_{2}) }
    {  \lambda_{1} \rho_{1,0} / m_{1} +   \lambda_{2} \rho_{2,0} / m_{2}} = 
\frac{1}{m^{*}}.
\end{equation}
As was mentioned earlier, the exact expressions for $\Omega_{1,2}(k)$ 
are invariant under the duality transformations of 
Section 3. It can be further checked that the approximate
expressions in (\ref{dispersion1}) retain this invariance.  
Thus, in particular, the expressions on the LHS's of (\ref{con1}) - 
(\ref{mstar}) are all invariant under the duality transformations of 
Section 3.

It is interesting to observe that for the special values of uniform densities configurations 
$\rho_{2,0}  =  \lambda_1\rho_{1,0}$, parameters $\lambda_1 \lambda_2 = \lambda_{12} = 1$, 
and masses $m_2 = m_1/\lambda_1$, the pieces of  both dispersions $\Omega_1(k)$ and $\Omega_2(k)$ 
which are proportional to $k^2$ vanish,  as should be evident from (\ref{dispersion1}) -  (\ref{mstar}). 
Thus, to order $k^2$,  $\Omega_2(k)$ vanishes identically, while $\Omega_1(k)$  attains the value of
the dispersion corresponding to the single-family Calogero model
\footnote{That is, $\Omega (k) = {1-\lambda\over 2m} k^2 + {\pi\lambda\rho_0\over m}|k|$.}
at its fermionic point $\lambda=1$. In other words,  at this particular point in parameter
space, our two-family Calogero model appears to be similar (but not entirely equivalent) to a system of 
free fermions.  The vanishing of $\Omega_2(k)$ at $k\neq 0$ is an indication of an instability at that point in 
parameter space. Evidently, one has to go beyond the quadratic approximation to resolve this problem. 

To summarize our discussion thus far, we have found that the low-momentum 
behavior of the two-family Calogero model is effectively that of an 
ordinary single-species Calogero model and a decoupled system of free massive
bosons. This conclusion will manifest itself again when we compute 
density-density  correlation functions in the next section.

Let us discuss now how to determine the effective one-family parameters 
$m, \lambda$ and $\rho$. Eqs. (\ref{con1}) and (\ref{con2}) comprise two 
relations among the three effective parameters $\rho, m $ and $\lambda\,.$ 
The third relation, needed to determine them unambiguously, arises from 
identifying the ground-state energy (\ref{e1}) with that of the 
single-species collective field \footnote{Clearly, the free quasi-particles 
described by $\Omega_2(k)$ cannot contribute to the ground state energy.}, 
namely, 
\beq\label{3rdcon}
\frac{{\pi}^{2}}{6 \lambda_{1} m_{1}} {(\lambda_{1} \rho_{1,0} + 
\lambda_{12} \rho_{2,0})}^{3}  =  \frac{{\pi}^{2} {\lambda}^{2}}{6 m} 
{\rho}^{3}\,.
\eeq
Note that although, strictly speaking, the first equality in the 
expression (\ref{e1}) for the ground state energy is based on the leading 
large-$N$ behavior of the effective potential (\ref{vofh}), the result
quoted there is, in fact, the {\em exact} value of the ground state energy, 
valid also for finite $N_{1,2}$. Indeed, the integrand appearing in that 
equation (namely, the LHS of (\ref{3rdcon})) is invariant under the duality transformations of 
Section 3, which hold also beyond leading order. Thus, we can treat 
(\ref{3rdcon}) on the same footing as (\ref{con1}) and (\ref{con2}), as far as
large-$N$ accuracy is concerned.

Recall that the couplings and masses are related according to 
(\ref{threebody}). Thus, eliminating, for example, $\lambda_2$ and 
$\lambda_{12}$ in terms of $\lambda_1$ and the ratio of masses, 
and then substituting these into (\ref{con1}), (\ref{con2}) and (\ref{3rdcon}),
it is possible to cast these three equations into forms from which 
elimination of $\lambda$ and $\rho$ in terms of $m$ leads to the 
simple relations 
\begin{equation} \label{con3}
 \frac{\lambda}{m^{2}} = \frac{\lambda_{1}}{{m_{1}}^{2}}\,,\quad\quad m\rho = 
m_1\rho_{1,0} + m_2\rho_{2,0}\,,
\end{equation}
while $m$ itself satisfies the quadratic equation
\beq\label{quadraticm}
\left(\rho_{1,0} + {m_2\over m_1}\rho_{2,0}\right)\,\left[
\lambda_1 \left({m\,{}\over m_1}\right)^2 - 1\right] + \left(
(1-\lambda_1)\rho_{1,0} + (1-\lambda_2)\rho_{2,0}\right) {m\,{}\over m_1} = 
0\,.
\eeq
(Note that Eqs.(\ref{con3}) and (\ref{quadraticm}) are invariant under the duality
transformations of Section 3, having been derived from duality-invariant 
equations.)

After solving (\ref{quadraticm}) for $m$ we can obtain $\lambda$ from 
(\ref{con3}). It is natural to conjecture that the quasi-particles of the effective 
one-family Calogero model have fractional statistics with statistical parameter 
$\lambda$. It will be interesting to prove this conjecture.

As an example for the discussion above, consider the case in which all couplings and 
masses are 
positive. Under these conditions, (\ref{quadraticm}) has two real solutions 
of opposite signs (which either remain invariant or are flipped by the duality
transformations). Let us choose the positive root as desired mass $m$. 
Substituting the latter into (\ref{con3}) we obtain $\lambda$ and $\rho$, 
which are also positive. (We do not present the corresponding expressions here.)

Finally, note that $\lambda=\lambda_1$, $m = m_1$, subjected to the 
constraints (\ref{democratic}) defining the SOHI, namely, the situation 
corresponding to (\ref{ftn1}), is also a solution of (\ref{con3}) and 
(\ref{quadraticm}). 


\section{Ground-state wave-functional and correlation functions}
Having completed diagonalization of $H_{coll}$ in (\ref{grrr}), we next
compute the corresponding collective-field ground-state wave functional. 
The ground state $ \; | 0 \rangle \; $
is defined by $ \; b_{1}(k) | 0 \rangle = b_{2}(k) | 0 \rangle = 0,  \; $ 
for every $ \; k $. Due to the orthogonal transformation (\ref{orttr}), the 
latter conditions are equivalent to $ \; a_{1}(k) | 0 \rangle = a_{2}(k) 
| 0 \rangle = 0,  \; $ or in real space, to 
\beq\label{realspace}
A_{1}(x) | 0 \rangle = A_{2}(x) | 0 \rangle = 0\,.
\eeq
We shall now concentrate on the latter representation. Following the 
definitions (\ref{grrr1}) and (\ref{grrr2}) we write (\ref{realspace}) as
\beqra\label{functionalde}
\left( -\partial_{x} \frac{\delta}{\delta \eta_{1}(x)} 
+ \frac{\lambda_{1} - 1}{2 \rho_{1,0}} \partial_{x} \eta_{1}(x) +
\lambda_{1} \pv \int dy \frac{\eta_{1}(y)}{x-y} + \lambda_{12} \pv \int dy 
\frac{\eta_{2}(y)}{x-y}  \right) \Psi_0[\eta_1,\eta_2] &=& 0\nonumber\\
{}\nonumber\\
\left( -\partial_{x} \frac{\delta}{\delta \eta_{2}(x)} 
+ \frac{\lambda_{2} - 1}{2 \rho_{2,0}} \partial_{x} \eta_{2}(x) +
\lambda_{2} \pv \int dy \frac{\eta_{2}(y)}{x-y} + \lambda_{12} \pv \int dy 
\frac{\eta_{1}(y)}{x-y}  \right) \Psi_0[\eta_1,\eta_2] &=& 0\,,\nonumber\\
{}
\eeqra
where $ \Psi_0[\eta_1,\eta_2] = \langle \{\eta_1,\eta_2\}| 0 \rangle$ is the 
wave functional we sought for. The solution of these linear 
and homogeneous coupled functional differential equations is evidently a 
Gaussian in the fluctuating fields $\eta_1$ and $\eta_2$, which we readily 
find as 
$$
\Psi_0[\eta_1,\eta_2] = {1\over \sqrt{\cz_0}}
\exp \int dx dy \left\{ \eta_{1}(x) 
\left( \frac{\lambda_{1} - 1}{4 \rho_{1,0}} \delta(x-y) + 
\frac{\lambda_{1}}{2} \ln|x-y| \right) \eta_{1}(y)\right. 
$$
\begin{equation} \label{functional}
\left. + \eta_{2}(x) \left( \frac{\lambda_{2} - 1}{4 \rho_{2,0}} \delta(x-y) +
\frac{\lambda_{2}}{2} \ln|x-y| \right) \eta_{2}(y) +  
\lambda_{12} \eta_{1}(x) \ln|x-y| \eta_{2}(y) \right\}\,,
\end{equation} 
where the normalization constant $\cz_0$ is fixed by the requirement that 
$\int\cd\eta_1\cd\eta_2 |\Psi_0[\eta_1,\eta_2]|^2 = 1\,.$
We may write (\ref{functional}) more compactly as 
\begin{equation}\label{functional1}
\Psi_0[\eta_1,\eta_2] = {1\over \sqrt{\cz_0}}
\exp \biggl( - \frac{1}{4} \int \int dx dy \sum_{a, b = 1}^{2}
   \eta_{a}(x) K_{ab} (x - y) \eta_{b}(y) \biggr),
\end{equation}
where
\beqra\label{kernel}
K_{11}(x) &=& \frac{1 - \lambda_{_1}}{\rho_{1,0}} \delta(x) -
 2\lambda_{1} \ln |x|\,,\nonumber\\
K_{22}(x) &=& \frac{1 - \lambda_{_2}}{\rho_{2,0}} \delta(x) -
 2\lambda_{2} \ln |x|\,,\nonumber\\
K_{12}(x) &=& K_{21}(x) = - 2\lambda_{12} \ln |x|\,.
\eeqra
By comparing (\ref{Acommutators}) and (\ref{kernel}) we obtain the simple 
relation 
\beq\label{AK}
[A_a(x), A_b^\dgg (y)] = \pa_x\pa_y K_{ab} (x-y)\,.
\eeq
Now that we have the collective-field vacuum wave-functional, we can calculate
the density-density correlation functions 
\beq\label{dosdos}
\langle 0 | \eta_a (x)\eta_b (y)| 0 \rangle =  {1\over \cz_0}
\int\cd\eta_1\cd\eta_2\, \eta_a (x)\eta_b (y)
\exp \biggl( - \frac{1}{2} \int \int dx dy \sum_{c, d = 1}^{2}
   \eta_{c}(x) K_{cd} (x - y) \eta_{d}(y) \biggr)
\eeq
by standard quantum field theoretic methods. We thus obtain  
\beq\label{dosdos1}
\langle 0 | \eta_a (x)\eta_b (y)| 0 \rangle =  K^{-1}_{ab} (x - y)\,, 
\eeq
where $ \; K^{-1}_{ab} (x - y)  \; $ denotes the inverse kernel
\begin{equation}
\int dz \sum_{c=1}^2 K_{ac} (x - z) K_{cb}^{-1} (z - y) = 
\delta_{ab}\delta(x - z)\,.
\end{equation}
Throughout the discussion from (\ref{functional}) to (\ref{dosdos1}) we have
assumed, of course,  that $K_{ab} (x - y)$ is positive definite.
Since the kernels $ \; K_{ab} (x - y) \; $ are translation-invariant, 
it is more appropriate to calculate the inverse kernels in
momentum space. Thus, by Fourier transforming both sides of (\ref{AK}), we 
find, using (\ref{Fourier}), (\ref{aofk}) and (\ref{adggofk}) - 
(\ref{omega}) that 
\beq\label{FourierK1}
K_{ab}(x) = \pv\int_{-\infty}^\infty\,dk\, e^{ikx}\,\tilde K_{ab}(k)\,,
\eeq
with 
\beq\label{FourierK}
\tilde K_{ab}(k) = \sqrt{{m_a\over \pi \rho_{a0}}
{m_b\over \pi \rho_{b0}}}{\om_{ab}(k)\over k^2}\,.
\eeq
Note that $\tilde K_{ab}(k)$ diverges like $1/|k|$ as $k\rightarrow 0$. 
Eq. (\ref{FourierK}) can be also verified by Fourier transforming all 
equations in (\ref{kernel}), by employing the Fourier integral 
\begin{equation}
\ln|x| = - \frac{1}{2} \pv \int_{-\infty}^\infty\, dk\, 
\frac{ e^{i kx}}{|k|}
\end{equation}
(and the standard Fourier representation of $\delta (x)$).
The desired inverse kernel is then found to be 
\beq\label{inverseK} 
K^{-1}_{ab} (x - y) = {1\over (2\pi)^2}\,\int\limits_{-\infty}^\infty\,dk
\, e^{ik(x-y)}\, \tilde K^{-1}_{ab} (k)\,,
\eeq
where $\tilde {\bf K}^{-1} (k) $ is the matrix inverse to (\ref{FourierK}). 
>From (\ref{FourierK}) and (\ref{commutator1}) - (\ref{omega}) we obtain the 
latter as 
\beq\label{inverseKk}
\tilde{\bf K}^{-1}(k) = 
{{2\pi\rho_{1,0}\rho_{2,0}\over\kappa}\over 1 + 
{(1-\lambda_1)(1-\lambda_2)\over 2\pi\kappa} |k|}\,\left(\begin{array}{cc}
{1-\lambda_2\over 2\pi\rho_{2,0}} |k| + \lambda_2  &  -\lambda_{12} \\
{} & {}\\
-\lambda_{12} & {1-\lambda_1\over 2\pi\rho_{1,0}} |k| + \lambda_1 
\end{array}
\right)\,,
\eeq
where 
\begin{equation}\label{kappa}
 \kappa =  (1 - \lambda_{1}) \lambda_{2} \rho_{2,0} + (1 - \lambda_{2}) 
\lambda_{1} \rho_{1,0}\,.
\end{equation}
 In obtaining (\ref{inverseKk}) we used 
\beq\label{detK}
\det\tilde{\bf K}(k) = {1\over 2\pi\rho_{1,0}\rho_{2,0}|k|} \left[\kappa 
+ {(1-\lambda_1)(1-\lambda_2)\over 2\pi} |k|\right]\,,
\eeq
where (\ref{threebody}) was invoked. 
By substituting (\ref{inverseKk}) in (\ref{inverseK}) we obtain the
integral representation 
\beqra\label{integralrepdos}
&& \langle 0 | \eta_a (x)\eta_b (0)| 0 \rangle  = K^{-1}_{ab} (x) = 
{1\over 2\pi} \left(\begin{array}{cc}
{\rho_{1,0}\over 1-\lambda_1}    &   0  \\
{} & {}\\
0  & {\rho_{2,0}\over 1-\lambda_2} 
\end{array}
\right)\,\delta(x) + 
\nonumber\\
{}\nonumber\\
&- &
2\left(\begin{array}{cc}
{\lambda_1\rho_{1,0}^2\over (1-\lambda_1)^2}    &   
{\lambda_{12}\rho_{1,0}\rho_{2,0}\over (1-\lambda_1)(1-\lambda_2)}  \\
{} & {}\\
{\lambda_{12}\rho_{1,0}\rho_{2,0}\over (1-\lambda_1)(1-\lambda_2)}  
& {\lambda_2\rho_{2,0}^2\over (1-\lambda_2)^2} 
\end{array}
\right)\,\int\limits_0^\infty\,dk\, 
{\cos (kx)\over k + { 2\pi\kappa\over(1-\lambda_1)(1-\lambda_2)}}
\eeqra
for the desired correlators (\ref{dosdos1}). The latter integral is well 
defined for 
\beq\label{q}
q = { 2\pi\kappa\over(1-\lambda_1)(1-\lambda_2)} = 2\pi\left(
{\lambda_1\rho_{1,0}\over 1-\lambda_1} + 
{\lambda_2\rho_{2,0}\over 1-\lambda_2}
\right)>0\,,
\eeq
in which case it can be expressed as a combination of the Sine-integral 
$\si (z) = -\int\limits_z^\infty\,dt (\sin t/t)$ and 
the Cosine-integral  $\ci (z) = -\int\limits_z^\infty\,dt (\cos t/t)$
\cite{abr}, leading to the final expression (for 
$x\neq 0$) 
\beq\label{dosdos2}
\langle 0 | \eta_a (x)\eta_b (0)| 0 \rangle = 2\left(\begin{array}{cc}
{\lambda_1\rho_{1,0}^2\over (1-\lambda_1)^2}    &   
{\lambda_{12}\rho_{1,0}\rho_{2,0}\over (1-\lambda_1)(1-\lambda_2)}  \\
{} & {}\\
{\lambda_{12}\rho_{1,0}\rho_{2,0}\over (1-\lambda_1)(1-\lambda_2)}  
& {\lambda_2\rho_{2,0}^2\over (1-\lambda_2)^2} 
\end{array}
\right)\,F(q|x|)\,,
\eeq
where
\beq\label{Fqx}
F(q|x|) =  \sin\, (q|x|) \,\si\, (q|x|) + \cos\, (q|x|)\,\ci \, (q|x|)\,.
\eeq
Eq. (\ref{dosdos2}) generalizes a similar result known for the 
one-family Calogero model \cite{Andric:1994su}.
The function $F(q|x|)$ has oscillatory behavior for small to moderate values
of its argument (as one would typically expect of correlation functions), 
and falls like $1/(qx)^2$ at large values. In fact, $F(q|x|)$ has a rather 
simple asymptotic expansion \cite{abr} \footnote{It is 
straightforward to obtain this asymptotic expansion by repeatedly 
integrating-by-parts the integral in (\ref{integralrepdos}).} in inverse 
powers of $(qx)^2$:
\beq\label{Fasymptotic}
F(q|x|) \sim \sum_{n=1}^\infty {(-1)^n\, (2n-1)!\over (qx)^{2n}}\,,
\quad\quad q|x|>>1\,. 
\eeq

In the case $q<0$ the integral representation (\ref{integralrepdos})
is ill-defined, due to the pole on the integration contour, however, 
the asymptotic series (\ref{Fasymptotic}) still makes sense, as the 
large-$|x|$ behavior of the correlation function comes from momentum 
values $|k| << |q|$. In either case, we can trust the asymptotic series
\beqra\label{asymptoticdos}
&& \langle 0 | \eta_a (x)\eta_b (0)| 0 \rangle  = K^{-1}_{ab} (x)\nonumber\\
{}\nonumber\\
&\sim &\!\!\!\!\!\!
\left(\begin{array}{cc}
{2\lambda_1\rho_{1,0}^2\over (1-\lambda_1)^2}    &   
{2\lambda_{12}\rho_{1,0}\rho_{2,0}\over (1-\lambda_1)(1-\lambda_2)}  \\
{} & {}\\
{2\lambda_{12}\rho_{1,0}\rho_{2,0}\over (1-\lambda_1)(1-\lambda_2)}  
& {2\lambda_2\rho_{2,0}^2\over (1-\lambda_2)^2} 
\end{array}
\right)
\sum_{n=1}^\infty (-1)^n\, (2n-1)!\left({(1-\lambda_1)(1-\lambda_2)
\over 2\pi\kappa x}\right)^{2n} \,,
\eeqra
valid for ${2\pi\kappa |x| \over   
(1-\lambda_1)(1-\lambda_2)} >> 1\,,$ to capture the leading 
long-distance behavior of the correlators. 

Close inspection of (\ref{dosdos2}) reveals that the coefficient matrix
there is of rank-1. Indeed, due to (\ref{threebody}) it can be written as 
${\bf w}{\bf w}^T$ with $w_a = {\sqrt{\lambda_a}\rho_{a,0}\over 1-
\lambda_a}\,.$ Thus, $\det\,\langle 0 | \eta_a (x)\eta_b (0)| 0 \rangle =0\,$
for all $ x\neq 0\,.$\footnote{Related to this is the fact that the matrix 
$\tilde{\bf K}^{-1}(0)$ exists, but due to (\ref{threebody}) it is not 
invertible.} More importantly, this means that only one combination 
of the density fluctuation fields $\eta_a (x)$ can have long-range 
correlations. In order to reveal the reason for this behavior, it is 
instructive to rewrite (\ref{functional1}) in momentum space. Starting
from the identity
$$\cs = \int \int dx dy\,\sum_{a, b}\eta_{a}(x) K_{ab} (x - y) 
\eta_{b}(y) = (2\pi)^2 \int\,dk\,\sum_{a, b}\tilde\eta_a(-k)
\tilde K_{ab}(k)\,\tilde\eta_b(k)$$  
we obtain, using (\ref{FourierK}), (\ref{diagonalization}) and $u_{ab}(-k) = 
u_{ab}(k)$ (see (\ref{phik})) that 
\beq\label{diagonalS}
\cs = (2\pi)^2 \int\,dk\,\sum_a\tilde\zeta_a(-k){\Omega_{aa}(k)\over k^2}\,
\tilde\zeta_a(k)\,,
\eeq
where 
\beq\label{zeta}
\tilde\zeta_a (k)  = \sum_b u_{ab}(k)\,\sqrt{m_b\over \pi\rho_{b,0}}\,\tilde
\eta_b(k)\,.
\eeq
Thus, (\ref{functional1}) factors into two {\em uncorrelated} Gaussian 
pieces, describing the vacuum fluctuations of the two fields 
$\zeta_a (x) = \int\, dk \, e^{ikx}\,\tilde\zeta_a (k)$.
>From the low-momentum behavior (\ref{dispersion1}) of $\Omega_1(k)$ and 
$\Omega_2(k)$, and from (\ref{con1}) - (\ref{mstar}), we conclude 
immediately that at long-distances, the vacuum fluctuations of the 
combination $\zeta_1 (x)$ coincide with those of a single-species Calogero
model with parameters $\lambda, m$ and $\rho$, whereas $\zeta_2(x)$ appears 
like a white-noise random field with ultra-local correlations proportional
to $\delta (x-y)$. All this, of course, is in full accordance with our
earlier conclusion that the two-family Calogero model is equivalent, in the 
large-$N$ limit and at long-distances, to two decoupled systems, one - an 
effective one-species Calogero model, and the other, a system of free massive
particles. 

As a simple application of this observation, consider the special two-family
systems living on the SOHI (\ref{democratic}) in parameter space. Let us
use our result (\ref{dosdos2}) to evaluate the correlation function of the 
effective one-family density (\ref{ftn1}). We shall denote the fluctuating 
part of $\rho(x)$ in (\ref{ftn1}) by $\eta (x) = \eta_1(x) - {1\over \lambda} 
\eta_2(x)$. Substituting the parameters corresponding to (\ref{democratic}) 
and (\ref{ftn1}) in the appropriate places, we obtain 
\beqra\label{specialcorrelator}
&&\langle 0 | \eta (x)\eta (0)| 0 \rangle =
\langle 0 | \left( \eta_{1}(x) - 
\frac{1}{\lambda}\eta_{2}(x) \right)
  \left( \eta_{1}(0) - \frac{1}{\lambda}\eta_{2}(0) \right)  | 0 \rangle
\nonumber\\{}\nonumber\\
&=&  \langle 0 | \eta_{1} (x)\eta_{1} (0)| 0 \rangle
 - \frac{2}{\lambda} \langle 0 | \eta_{1} (x)\eta_{2} (0)| 0 \rangle
 + \frac{1}{{\lambda}^{2}} \langle 0 | \eta_{2} (x)\eta_{2} (0)| 0
 \rangle
\nonumber\\{}\nonumber\\
 &=& {2\lambda\over (1-\lambda)^2} \left(\rho_{1,0} - 
{1\over\lambda}\rho_{2,0}\right)^2 F(q|x|)\,,
\eeqra
with 
\beq\label{specialq}
q = {2\pi \lambda\over\lambda -1} \left(\rho_{1,0} - 
{1\over\lambda}\rho_{2,0}\right)\,.
\eeq
>From (\ref{Fasymptotic}) we obtain the leading long-distance asymptotic 
behavior of this correlator simply as 
\beq\label{asymptoticdosdos}
\langle 0 | \eta (x)\eta (0)| 0 \rangle = 
-\frac{1}{2 {\pi}^{2} \lambda}  \frac{1}{x^{2}},
\end{equation}
which does not depend on the particle density at all, in accordance
with the one-family correlation function of \cite{Sen:1997qt, ha}.

\subsection{The low-lying collective excitations}
We close this section by analyzing of the low-lying collective excitations
above the vacuum. These excitations are the pseudo-particles created by 
$b_{1,2}^\dgg (k)$ acting on the vacuum. The corresponding states are 
\beq\label{excited}
b_a^\dgg (k) | 0 \rangle = {1\over 2\sqrt{\pi}}\,\int\limits_{-\infty}^
\infty\, dx\, e^{ikx} \sum_b u_{ab} (k) \sqrt{{\rho_{b0}\over m_b}}\,
A_b^\dgg (x) | 0 \rangle\,,
\eeq
where we used (\ref{orttr}), (\ref{Fourier}), (\ref{grrr1}), (\ref{grrr2}) 
and the fact that the $u_{ab}$ are real.  Next, we observe from (\ref{grrr1}),
(\ref{grrr2}) and (\ref{functional}) that 
\beq\label{Adggvac}
A_a^\dgg (x)  | 0 \rangle = -2i\left( 
\frac{\lambda_{a} - 1}{2} \frac{\partial_{x} \eta_{a}}{\rho_{a0}} +
 \lambda_{a} \pv \int \frac{ dy \eta_{a}(y)}{x - y}  +
 \lambda_{12} \pv \int \frac{dy \eta_{\bar a}(y)}{x - y} \right)
| 0 \rangle 
\eeq
where $\bar a$ is the index obtained from $a$ by permutation, i.e., 
$\bar 1 = 2, \bar 2 = 1$. By substituting (\ref{Adggvac}) in (\ref{excited}), 
and using (\ref{pv}), we then obtain, after some algebra, 
\beq\label{excited1}
b_a^\dgg (k) | 0 \rangle = {1\over \sqrt{\pi}\,k}\,\int\limits_{-\infty}^
\infty\, dx\, e^{ikx} \sum_b u_{ab} \left(\sqrt{{m_b\over \rho_{b0}}}\,
\om_{bb} \,\eta_b(x) + \sqrt{{m_{\bar b}\over \rho_{{\bar b}0}}}\,
\om_{b{\bar b}} \,\eta_{\bar b}(x)\right)| 0 \rangle\,,
\eeq
where we also used (\ref{commutator1}) - (\ref{commutator}) on the way. 
Performing summation over the index $b$ and using (\ref{diagonalization}), 
which is uquivalent to the relations $u_{aa}\,\om_{aa} + u_{a{\bar a}}
\om_{{\bar a}a} = \Om_a u_{aa}$ and $u_{aa}\om_{a{\bar a}} + 
u_{a{\bar a}}\om_{{\bar a}{\bar a}} = \Om_a u_{a{\bar a}}$\,,
we may simplify (\ref{excited1}) further into 
\beq\label{excited2}
b_a^\dgg (k) | 0 \rangle = {\Om_a (k)\over k}\,\int\limits_{-\infty}^
\infty\, dx\, e^{ikx} 
\left(\sqrt{{m_a\over \pi\rho_{a0}}} u_{aa}\,\eta_a(x)
+\sqrt{{m_{\bar a}\over \pi\rho_{{\bar a}0}}} u_{a{\bar a}} \,\eta_{\bar a}(x)
\right)
| 0 \rangle\,.
\eeq
Using the definitions of $u_{ab}$ given above (\ref{rotation}) we thus obtain
\beqra\label{b1b2}
b_{1}^{\dagger} (k) | 0 \rangle &\propto &
\sqrt{\frac{m_{1} \rho_{2,0}}{m_{2} \rho_{1,0}}} \,\cot \phi 
\int\limits_{-\infty}^\infty\, dx \,e^{ikx} \eta_{1}(x) | 0 \rangle + 
\int\limits_{-\infty}^\infty\, dx \,e^{ikx} \eta_{2}(x) | 0 \rangle\nonumber\\
{}\nonumber\\
b_{2}^{\dagger} (k) | 0 \rangle &\propto &
- \sqrt{\frac{m_{1} \rho_{2,0}}{m_{2} \rho_{1,0}}} \,\tan \phi 
\int\limits_{-\infty}^\infty\, dx \,e^{ikx} \eta_{1}(x) | 0 \rangle + 
\int\limits_{-\infty}^\infty\, dx \,e^{ikx} \eta_{2}(x) | 0 \rangle\,.
\eeqra
Let us now compute $\cot\phi$. Substituting (\ref{rotation}) in the identity 
$\cot\phi = \cot 2\phi + \sqrt{1+\cot^2 2\phi}$ and then using 
(\ref{commutator1}) - (\ref{commutator}) we obtain 
\beq\label{cot}
\cot\phi = {\om_1-\om_2 + \Om_1-\Om_2\over 2\om_{12}}\,,
\eeq
which upon expansion to leading order in $|k|$ yields
\beq\label{cot1}
\cot\phi \simeq \sqrt{m_2\rho_{1,0}\over m_1\rho_{2,0}}\,
{\lambda_1\over \lambda_{12}}\left(
1+ {1\over 2\pi} {{1-\lambda_1\over m_1}-{1-\lambda_2\over m_2}
\over 
{\lambda_1\rho_{1,0}\over m_1} + {\lambda_2\rho_{2,0}\over m_2}}\,|k|
\right)\,.
\eeq
Finally, returning to the original configuration
space coordinates
\beq
\int dx e^{ikx} \eta_{1}(x) \rightarrow \sum_{i=1}^{N_{1}} e^{ikx_{i}}\,,
\quad\quad 
\int dx e^{ikx} \eta_{2}(x) \rightarrow  \sum_{\alpha=1}^{N_{2}} 
e^{ikx_{\alpha}}\,,
\eeq
and substituting it and (\ref{cot1}) in (\ref{b1b2}) we find
\begin{equation}\label{bexcitation1}
b_{1}^{\dagger} (k) | 0 \rangle \propto  \frac{m_{1}}{m_{2}}
  \left( 1 + \frac{1}{2 \pi}
 \frac{ \frac{1 - \lambda_{1}}{m_{1}} - \frac{1 - \lambda_{2}}{m_{2}}}
   {\frac{\lambda_{1} \rho_{1,0} }{m_{1}} + \frac{\lambda_{2}
  \rho_{2,0} }{m_{2}}}  |k| \right) \sum_{i=1}^{N_{1}} e^{ikx_{i}} | 0\rangle
  + \sum_{\alpha=1}^{N_{2}} e^{ikx_{\alpha}} | 0 \rangle
\end{equation}
and 
\begin{equation} \label{bexcitation2}
  b_{2}^{\dagger} (k) | 0 \rangle \propto - \frac{\rho_{2,0}}{\rho_{1,0}}
  \left( 1 - \frac{1}{2 \pi}
 \frac{ \frac{1 - \lambda_{1}}{m_{1}} - \frac{1 - \lambda_{2}}{m_{2}}}
   {\frac{\lambda_{1} \rho_{1,0} }{m_{1}} + \frac{\lambda_{2}
  \rho_{2,0} }{m_{2}}}  |k| \right) \sum_{i=1}^{N_{1}} e^{ikx_{i}} | 0\rangle
  + \sum_{\alpha=1}^{N_{2}} e^{ikx_{\alpha}} | 0 \rangle\,,
\end{equation}
where we used $\lambda_1/\lambda_{12} = m_1/m_2$ from (\ref{threebody}). 
Our low-energy excitations are in fact long-wavelength density oscillations
(phonons) in which the velocities of all particles point in the same
direction, either positive or negative, depending on the sign of $ \; k.$

Comparing the excitation (\ref{bexcitation1}) with that obtained in \cite{Sen:1995zj}, we
see that our relative weight approaches the relative weight 
$ \; m_{1}/m_{2} \; $ from \cite{Sen:1995zj}, but only in the small $ \; k \; $
limit.
The second excitation (\ref{bexcitation2}), which describes the free-particle motion,
does not have any counterpart in Sen's approach \cite{Sen:1995zj}.


\section{The multi-species generalization}
Let us now apply the methods we used in our analysis of the two-family 
Calogero model to investigate its $F$-family generalization. 
Thus, consider a collection of $ \; F $ species of particles. 
The $ \; a- $ th family contains $ \; N_{a} \; $
particles of mass $ \; m_{a}\quad (a = 1,2,\ldots F) \; , $ which 
interact among themselves with coupling constant $ \; {\lambda}_{a}\equiv 
{\lambda}_{aa} . $ The inter-family mutual interaction 
strengths are $ \; {\lambda}_{ab} = \lambda_{ba}\quad (a\neq b)\,.\; $ 
All couplings satisfy the constraint
\begin{equation} \label{3body}
 \frac{\lambda_{ab}}{m_{a} m_{a}} = c\,,
\end{equation}
where the same constant $c$ holds for all pairs of indices. This condition 
generalizes (\ref{threebody}) and guarantees that there be no 
three-body interactions \cite{Meljanac:2003jj, Meljanacstojic}. It follows from (\ref{3body}) that 
\beq\label{lambdaab}
\la_a\la_b = \la_{ab}^2\,,
\eeq
similarly to (\ref{lambda12}). 
We take the limit in which all particle numbers $N_a$ tend to infinity at 
the same rate. The collective-field Hamiltonian of the $F-$family Calogero 
model is given by
\begin{equation} \label{Fham}
 H_{coll} = \sum_{a = 1}^{F}
\frac{1}{2 m_{a}} \int dx \, \left(\pax\pi_a(x)\rho_a (x)\pax\pi_a(x)
+ \rho_{a}(x) \,{\left( \frac{\lambda_{a} - 1}{2} \frac{\partial_{x} \rho_{a}}
{\rho_{a}} +
\sum_{b = 1}^{F} \lambda_{ab} \pv \int \frac{ dy \rho_{b}(y)}{x - y}
\right)}^{2} \right),
\end{equation}
up to singular boundary-contributions (compare to (\ref{colham})). 
The collective density fields and their conjugate momenta obey canonical 
equal-time commutation relations analogous to (\ref{canonical}). 
The $a-$th density field is 
subjected, of course, to the normalization condition
$\int\,dx \rho_a(x) = N_a$.

The ground-state energy and density 
distributions can be found, as in the 2-family case, by varying the effective 
potential corresponding to (\ref{Fham}) with respect to all 
$\; {\rho}_{a}\,. $ We thus obtain the uniform ground-state densities 
\begin{equation} \label{unif}
 {\rho}_{a,0} = \frac{N_{a}}{L}\,,
\end{equation}
where $L$ is the length of the large confining box, and the ground-state 
energy 
\begin{equation} \label{gren}
 E_{0} = \frac{{\pi}^{2} {c}^{2}}{6 {L}^{2}} 
\left( \sum_{a = 1}^{F} m_{a} N_{a}\right)^{3} = 
{\pi^2\over 6\lambda_1 m_1 L^2}\left(\sum_{a=1}^F \lambda_{1a} N_a\right)^3.
\end{equation}
In the last equation, which is a direct generalization of (\ref{e1}), 
we used (\ref{3body}). 

Similarly to our conventions in the $F=2$ case, we shall take 
all $\lambda_a > 0 $, and anticipating the possibility of negative masses
and densities, we shall also set ${\rm sign}\,m_a =  {\rm sign}\,\rho_{a,0}$
for all families.

We now expand around the uniform configurations (\ref{unif})
\begin{equation} \label{exp}
   \rho_{a} (x) = \rho_{a,0} + \eta_{a} (x),
\end{equation}
and introduce the $F$ operators $A_c(x)$ analogous to (\ref{grrr1}) and 
(\ref{grrr2}). The Fourier transforms $a_c(k)$ of these operators, and their
hermtian conjugates, defined as in (\ref{aofk}) and (\ref{adggofk}), 
obey the commutation relations $[a_a(k), a_b^\dgg (k')] = \omega_{ab} (k) 
\delta(k - k')\,,$ where the $F\times F$ real symmetric matrix  
\begin{equation}\label{omegaF}
  {\omega}_{ab}(k) =  \frac{1 - \lambda_{a}}{2 m_{a}} k^{2}
  {\delta}_{ab} +
  \lambda_{ab} \pi \sqrt{\frac{\rho_{a,0} \rho_{b,0}}{m_{a} m_{b}}} |k|  
\end{equation}
is an immediate generalization of (\ref{omega}).

Let us now diagonalize $\omega_{ab}$ and obtain its eigenvalues $\Omega_a(k)$.
We shall contend ourselves with the low-momentum behavior of these
dispersions, and compute them to order $k^2$. 
Details of this computation are given in Appendix B. We find that one of 
these eigenvalues is given approximately by
\begin{equation} \label{eigenvalues}
  {\Omega}_{1}(k) = \frac{\sum_{a = 1}^{F} \frac{(1 - \lambda_{a})
  \lambda_{a} \rho_{a,0} }{{m_{a}}^{2}} } {\sum_{i = a}^{F}
 \frac{\lambda_{a} \rho_{a,0} }{m_{a}}} \frac{k^{2}}{2}                  
   + \pi \sum_{a = 1}^{F} \frac{\lambda_{a}\rho_{a,0}}{m_{a}} |k|\,.  
\end{equation}
Since in our conventions $\lambda_a >0 $ and ${\rm sign}\,m_a = {\rm sign}\,
\rho_{a,0}$, the coefficient of $|k|$ in the last term is manifestly positive.
The remaining $F-1$ eigenvalues 
$\; {\Omega}_{2}(k) \sim {\Omega}_{3}(k) \sim \ldots \sim
{\Omega}_{F}(k) \sim k^{2}. $ Thus, as in the $F=2$ case, the first 
dispersion $\;{\Omega}_{1}(k) \; $ looks effectively like that of a 
one-family Calogero model. The remaining $F-1$ dispersions correspond 
evidently to free massive bosonic quasi-particles, decoupled from each
other and from the Calogero-like mode.

Let us now compare our Calogero-like dispersion law (\ref{eigenvalues}) 
with that given by Eq.(58) of \cite{Sen:1995zj}, which was derived to leading 
order in $\; \frac{1}{L}. $
The terms linear in momentum $\; k \; $ agree, as can be easily seen by
virtue of the relation
\begin{equation}\label{vf}
 v_{F} = \pi \sum_{a} \frac{\lambda_{a} \rho_{a,0}}{m_{a}},
\end{equation}
where $\; v_{F}  \; $ denotes the ''Fermi'' velocity in Sen's
approach. We observe that the $\; k^{2} \; $ correction in our
dispersion (\ref{eigenvalues}) arises from both the linear and the quadratic 
differential operators in the Hamiltonian (\ref{Fham}). Taken together, they 
give a contribution of the typical weight $\; \frac{\lambda_{a} - 1}{2}. $
These terms have been completely ignored in Sen's corresponding
dispersion relation mentioned above, because they did not seem to be
analytically computable in that approach. In this respect, our method
seems to do better than the generalized Thomas-Fermi method of 
\cite{Sen:1995zj}.

\subsection{Duality transformations and symmetries of the $F$-family model}
The duality symmetries of the two-family model, discussed in Section 3, 
have a natural generalization in the $F$-family model. 
The main idea in searching for these duality symmetries in the generic 
case is to find all classes of physically equivalent $ \; F- $ family models. 
Here, the eqivalence of two systems in a given class means that despite the 
fact that these systems are defined in terms of different sets of collective 
fields, momenta and parameters, their collective Hamiltonians (\ref{Fham})
coincide.

As our first step in classifying all these duality symmetries, it
is easy to check that the Hamiltonian (\ref{Fham}) is invariant under the 
following set of elementary {\em duality} transformations, associated with a 
prescribed family index $a$:
$$
 \tilde{\lambda}_{a} = \frac{1}{{\lambda}_{a}}; \;\;\;\;
 \tilde{\lambda}_{ab} = - \frac{{\lambda}_{ab}}{{\lambda}_{a}}, \;\;
 \mbox{for} \;\; b \neq a;  \;\;\;\; 
\tilde{\lambda}_{cd} = {\lambda}_{cd}, \;\;
 \mbox{for} \;\; c,d  \neq a;
$$
$$
 \tilde{m}_{a} = - \frac{{m}_{a}}{{\lambda}_{a}};  \;\;\;\; 
\tilde{m}_{b} = m_{b}, \;\;  \mbox{for} \;\; b \neq a;
$$
$$
 \tilde{\rho}_{a} = - {\lambda}_{a}{\rho}_{a}; \;\;\;\;
  \tilde{\rho}_{b} = {\rho}_{b}, \;\; \mbox{for} \;\; b \neq a;
$$
\begin{equation} \label{duality}
 \tilde{\pi}_{a} = - \frac{{\pi}_{a}}{{\lambda}_{a}};  \;\;\;\; 
\tilde{\pi}_{b} = {\pi}_{b}, \;\; \mbox{for} \;\; b \neq a.
\end{equation}
Let us denote all these transformations collectively as $T_a\,,~~a=1,\ldots 
F\,.$ As can be readily seen, $T_a$ acts non-trivially only on the 
$ \; a - $ th family.  Clearly,  $T_a^2 = I\,,$ where $I$ is the identity 
transformation. Therefore, $T_a$ and $I$ form a $Z\!\!\! Z_2$ group of 
of symmetries of (\ref{Fham}). It is possible to compose any pair of such 
elementary transformations, corresponding to different family indices. 
Moreover, this composition is obviously commutative. In this way 
we can generate the more complicated duality symmetries 
$T_aT_b = T_bT_a$ of (\ref{Fham}). Continuing this process, we can generate 
all transformations involving 
triplets of different family indices $T_aT_bT_c$ all the way up to 
the maximally possible composition $T_1T_2\cdots T_F$. In this way, the 
elementary transformations $T_a$, together with the identity $I$ generate 
an Abelian group of order $2^F$, which exhaust all possible duality 
symmetries of (\ref{Fham}). Evidently, each element in this group is of order
2. We can readily identify this group is as $\otimes^F Z\!\!\! Z_2\,.$

Within this general
picture, we can interpret the two-family duality transformations
(\ref{duality6}), (\ref{duality7}) as an elementary transformation on the 
second family, while the transformations (\ref{duality1}),(\ref{duality2}) 
may be understood as the  simultaneous action of the two elementary 
transformations on the first and the second family.

All physical properties of the $F$-family system should remain invariant
under the duality symmetries of (\ref{Fham}). In particular, the $F$ 
dispersions $\Omega_a(k)$ should be duality-invariants. This can be proved 
in a manner similar to the $F=2$ case, namely by showing that the effect of 
a duality transformation on the matrix ${\bf\om}$ in (\ref{omegaF}) is 
simply a conjugation by an orthogonal matrix.\footnote{See the discussions 
following (\ref{omega}) and (\ref{Omega}).} It is enough to prove this 
conjugation for an elementary transformation (\ref{duality}): It can be easily 
checked that the elementary duality transformation $T_a$ leaves $\om_{aa}, 
\om_{bb}$ and $\om_{cd}$ ($b,c,d,\neq a $) invariant, while $\om_{ab} = 
\om_{ba}$ flips sign. All this can be summarized by saying that
under $T_a$, ${\bf\om}\rightarrow \tau_a {\bf\om} \tau_a$, where the 
diagonal orthogonal matrix $\tau_a$ is obtained from the unit matrix 
by flipping the sign of its $a$-th diagonal entry. Thus, all the 
dispersions $\Omega_a(k)$ are invariant under the duality transformations
of (\ref{Fham}). In particular, all duality transformations preserve 
positivity of the coefficient of $|k|$ in the second term in 
(\ref{eigenvalues}).


\subsubsection{Special $F$-family Calogero models which resemble  
single-family Calogero models}
In Section 3.1 we applied the duality group to identify a special subset
of two-family models, with apparently different families, which nevertheless
resemble, in almost all respects, a Calogero system composed of 
identical particles. Those are the two-family systems living on the 
so-called ``surface of hidden identity'' (SOHI) (\ref{democratic}) in 
parameter space. Not surprisingly, this construction can be generalized to 
the $F$-family case, as we now show. 

In order to find special multi-family Calogero models, which resemble 
a single-species Calogero model with some effective particle density 
$ \; \rho, \; $ mass $ \; m \; $ and interaction strength $ \; \lambda\,, $
let us split the latter effective one-family system into $ \; F \; $ 
different sub-families whose particles share the same mass  
$ \; m, $ the same coupling constant $ \; \lambda \; $ and
the same mutual interaction strength $ \; {\lambda}_{ab} = \lambda. $ 
This splitting means of course that we no longer consider totally symmetric 
many-body wave functions, and replace them by wave functions symmetric 
separately in the coordinates of the particles of each family. Let 
$\rho_a$ denote the density of $a$-th family particles. Therefore, 
$\rho = \sum_a \rho_a$. We shall refer to the $F$-family system thus 
obtained as maximally degenerate (MD). 

Let us consider our MD $F$-family model as the image under one
of the elementary duality transformations $T_k$ in (\ref{duality}) for 
some $1\leq  k \leq F$. Since $T_k^2 = I$, its origin must have been 
the less-degenerate $F$-family Calogero system with parameters 
\beqra\label{parameters1}
\lambda_k = {1\over\lambda}\,,\quad \lambda_{ak} &=& \lambda_{ka} = -1\,,
\quad \lambda_a = \lambda_{ab} = \lambda\quad {\rm for} \quad 
a,b\neq k\,;\nonumber\\ 
m_k  &=&  -{m\over\lambda}\,,\quad m_a = m \quad {\rm for} \quad
a\neq k\,;
\eeqra
partial densities
\beq\label{densities1}
-{\rho_k\over\lambda} \,,\quad \rho_a\quad {\rm for} \quad a\neq k\,,
\eeq
and total density 
\beq\label{total1}
\rho (x) = \sum_{a\neq k} \rho_a(x) - {1\over\lambda}\rho_k (x) \,.
\eeq
To reverse the argument, the $k$-th family in the less degenerate 
$F$-family system defined by (\ref{parameters1}) - 
(\ref{densities1}) is singled out, yet the latter is dual to the 
MD $F$-family system, and therefore resembles a single family 
Calogero model. We can now apply another element $T_l$ ($l\neq k$) of 
(\ref{duality}) on the less degenerate $F$-family system, and 
obtain an even lesser-degenerate $F$-family system, now with two 
families singled out. Yet, the latter system is stil dual to the MD system. 
We can clearly continue this process thus generating all images of the 
MD $F$-family system under all $2^F-1$ non-unit elemenets of the duality 
group. Each of these images comprises a special $F$-family system, whose 
families are not all identical, which nevertheless resembles a single family 
Calogero model. Just to give a nontrivial example, consider the image 
of the MD $F$-family system under the combined duality transformation 
$T_{k+1}T_{k+2}\cdots T_F$. This image is the $F$-family Calogero 
system with masses $ \; m_{a}, $ coupling constants $ \; {\lambda}_{a} \; $ 
and mutual interaction strengths $ \; {\lambda}_{ab} $ given by 
$$
 {\lambda}_{1} = {\lambda}_{2} = ... = {\lambda}_{k} = \lambda;
$$
$$
 m_{1} = m_{2} = ... = m_{k} = m;
$$
$$
{\lambda}_{ab} = \lambda, \;\; \mbox{for} \;\; a,b \leq k; 
  \;\;\;\; {\lambda}_{k + 1} = {\lambda}_{k + 2} = ... = {\lambda}_{F}
  = \frac{1}{\lambda};
$$
$$
 m_{k + 1} = m_{k + 2} = ... = m_{F} = -\frac{m}{\lambda};
$$
\begin{equation} 
 {\lambda}_{ab} = \frac{1}{\lambda}, \;\; \mbox{for} \;\; a,b > k; \;\;\;\; 
 {\lambda}_{ab} = -1, \;\; \mbox{for} \;\; a \leq k \;\; \mbox{and}
 \;\;  b > k.
\end{equation}
The effective one-family collective-field is obviously given by
\begin{equation}
 \rho(x) = \sum_{a = 1}^{k} \rho_{a}(x) - \frac{1}{\lambda} 
\sum_{a =  k + 1}^{F} \rho_{a}(x).
\end{equation}
For $F=2$ this result clearly coincides with (\ref{ftn1}).


\section{Concluding remarks}

We conclude this paper with a few comments on some recent papers and also 
on another set of duality transformations which were discussed in the literature.

Sergeev and Veselov \cite{sergeev1} constructed
supersymmetric extensions of the Calogero-Sutherland model which actually
correspond to our two-family Calogero model with $ \; \lambda_{1}
\lambda_{2} = 1, \; \lambda_{12} = -1 \; $ and $ \; m_{1} m_{2} < 0! $
They gave solutions in terms of deformed Jack polynomials. In a
recent paper, Kohler and Guhr \cite{guhr} introduced a supersymmetric
generalization of the Calogero-Sutherland model. Their construction is
based on Jacobians for the radial coordinates on certain
superspaces. This approach allowed them to explicitly construct the
solutions in terms of recursion formulae for a non-trivial 
$ \; ( \lambda_{1} \lambda_{2} = 1) \; $ one-parameter subspace in the 
$ \; ( \lambda_{1}, \lambda_{2}) \; $ plane.
The underlying model involves two kinds of interacting particles, one
with the positive and the other one with the negative mass. Needless to say,
this again corresponds to our two-family Calogero
 model with $ \; \lambda_{12} = -1. $
It is interesting to observe that the authors of Refs.\cite{sergeev1}-\cite{ Guhr:2004ff} were
probably unaware of the constraints (\ref{threebody}). Namely, in their approaches,
these constraints remain hidden, but still present, as can be easily
checked by direct substitutions. Consequently, the two types of models 
discussed in \cite{sergeev1}-\cite{ Guhr:2004ff}, share the very same parametric
structure, which enables one to transform them to the one-family
Calogero model. This connection then guarantees their exact
integrability. Although our collective-field approach is applicable
only to the multi-species Calogero system with an infinitely large
number of particles within each family, we believe that our findings
shed some light on the problem of their exact integrability in
general.

Let us finally remark on an additional {\em approximate} duality invariance
of the multi-species model (\ref{Fham}), known in the literature
\cite{andrjur, andrjur1}.  For simplicity, we
illustrate it on the two-family Hamiltonian (\ref{colham}), in which case 
this approximate duality symmetry involves special 
two-family systems, in which $\lambda_2 = {1\over \lambda_1}$ but 
$\lambda_{12} = +1$, as opposed to (\ref{democratic}). 
This duality symmetry is destroyed by subleading terms in (\ref{colham})
as we now explain.

Estimating the $\; \frac{1}{N} \; $
 dependence of the terms in the effective potential
(25), we see that the $ \; \partial_{x} ln \rho_{1} \; $ and 
$ \; \partial_{x} ln \rho_{2} \; $ terms are down by
$ \; \frac{1}{N_{1}} \; $  and by
$ \; \frac{1}{N_{2}}, \; $ respectively, when compared with the Hilbert-transform
terms. The effective potential to this order is then 
$$
 V =  \frac{1}{2 m_{1}} \int dx \rho_{1}(x) {\left( 
 \lambda_{1} \pv \int \frac{ dy \rho_{1}(y)}{x - y}  + \lambda_{12}
 \pv \int \frac{dy \rho_{2}(y)}{x - y} \right)}^{2}
$$
\begin{equation}
 + \frac{1}{2 m_{2}} \int dx  \rho_{2}(x) {\left( 
   \lambda_{2} \pv \int \frac{ dy  \rho_{2}(y)}{x - y}
  + \lambda_{12} \pv \int \frac{dy \rho_{1}(y)}{x - y} \right)}^{2}. 
\end{equation}
One can verify that the above potential is invariant under the
following set of transformations of the parameters:
$$
 \tilde{\lambda}_{1} = {\lambda}_{1}; \;\;\;\;
 \tilde{\lambda}_{2} = \frac{1}{{\lambda}_{2}}; \;\;\;\;
\tilde{m}_{1} = m_{1}, \;\;\;\;  \tilde{m}_{2} = \pm
 \frac{m_{2}}{\lambda_{2}}; \;\;\;\;
\tilde{\lambda}_{12} = \pm \frac{{\lambda}_{12}}{{\lambda}_{2}}
$$
and of the densities
\begin{equation}
 \tilde{\rho}_{1} = {\rho}_{2}; \;\;\;\;
  \tilde{\rho}_{2} = \pm  {\lambda_{2}} {\rho}_{2}.
\end{equation}
Arguing along the same lines as before, we can show that the
two-family Calogero model is equivalent to the one-family Calogero
model if the following conditions are satisfied:
$$
 {\lambda}_{1} = \lambda; \;\;\;\;
 {\lambda}_{2} = \frac{1}{\lambda}; \;\;\;\; 
{m}_{1} = m, \;\;\;\;  {m}_{2} = \pm
 \frac{m}{\lambda}; \;\;\;\;
{\lambda}_{12} = \pm 1,
$$
where $ \; \lambda \; $ denotes the coupling constant and $ \; m \; $
the mass of the effective one-family model. The effective one-family
collective density is given by
\begin{equation}
 \rho(x) = \rho_{1} (x) \pm \frac{1}{\lambda} \rho_{2} (x).
\end{equation}
We stress that the above picture is valid in the leading approximation
only. Consequently, this description is inevitably destroyed by the
next-to-leading-order terms stemming, for instance, from the quantum
collective-field fluctuations. In contrast, the various dualities 
$T_1, T_2, T_{12}$ (and, of course, 
the indentity $I$), discussed in Section 3, are {\em exact} symmetries of
the collective Hamiltonian (\ref{colham}). Consequently, the 
dual equivalence of any two-family system on the SOHI 
(\ref{democratic}) to a two-family system of identical families given by
(\ref{ftn1}) or (\ref{ftn2}) is also an exact property of (\ref{colham}).

\bigskip
\bigskip

\pagebreak

%


\newpage
\setcounter{equation}{0}
\setcounter{section}{0}
\renewcommand{\theequation}{A.\arabic{equation}}
\renewcommand{\thesection}{Appendix A:}
\section{An alternative derivation of (\ref{vofh})}
\vskip 5mm
\setcounter{section}{0}
\renewcommand{\thesection}{A}
For completeness, we present here an alternative derivation of the 
expression (\ref{vofh}) for the collective effective potential.
To this end, it is useful to introduce the wieghted density 
\beq\label{rho}
\rho(x) = \sqrt{\lambda_1} \rho_1(x) + \sqrt{\lambda_2}\rho_2 (x)\,,
\eeq
and consider the resolvent 
\beq\label{G}
G(z) = \int {dy \rho(y)\over z-y}
\eeq
associated with it, in which $z$ is a complex variable. As in the single 
family case, the particles are expected to condense in the 
ground state in a large but finite segment of some length $L$ - the support 
of $\rho (x)$. The resolvent $G(z)$ is therefore analytic 
in the complex plane, save for a cut along that segment, which lies on the 
real axis. Consequently, as usual, we have 
\beq\label{G1}
G(x\mp i0) = F(x) \pm i\pi\rho(x) \,
\eeq
where 
\beq\label{F}
F(x)  =  \pv \int \frac{ dy \rho (y)}{x - y}\,.
\eeq
The combinations of principal part integrals in the first two lines in 
(\ref{potential1}) are evidently proportional to $F(x)$. 
This fact, together with (\ref{lambda12}) can be used to rewrite 
(\ref{potential1}) as 
\beqra\label{potential2}
V &=&  {(\lambda_1-1)^2\over 8 m_1} \int dx {(\partial_{x} \rho_{1})^2\over 
\rho_{1}} + 
{(\lambda_2-1)^2\over 8 m_2} \int dx {(\partial_{x} \rho_{2})^2\over \rho_{2}}
\nonumber\\
&+& {\sqrt{\lambda_1}\over m_1}\int dx\, F(x)\, \partial_x 
\left({\lambda_1-1\over 2}\rho_1 + 
{\lambda_2-1\over 2}\rho_2\right)\nonumber\\
&+& {\sqrt{\lambda_1}\over 2m_1}\int dx\, \rho (x)\, F^2(x)\nonumber\\
&+& \mu_{1} \left(N_{1} - \int dx \rho_{1}(x) \right) + \mu_{2} \left(N_{2} - 
\int dx \rho_{2}(x)\right)\,.
\eeqra
The term $\rho (x)\, F^2(x) $ in the third line in (\ref{potential2}) appears 
to be a trilocal cubic functional of the densities. 
In fact, it can be brought to local form by a standard trick \cite{sakita}, 
based on the analytic structure of $G(z)$ and its asymptotic behavior 
\beq\label{Gasympt}
G(z)\sim {N_1\sqrt{\lambda_1} + N_2\sqrt{\lambda_2}\over z}
\eeq
as $z\rightarrow\infty$. Note that the fact that $\rho_1(x)$ and $\rho_2(x)$ 
have compact support is necessary for (\ref{Gasympt}) to hold. 
It follows from (\ref{Gasympt}) that 
$\oint_{C_\infty} dz \,G^3(z)  = 0 $, where $C_\infty$ is a circle
around the point at infinity. Then, shrink  $C_\infty$ to wrap around the cut 
to obtain $ 2i \int dx\, \Im G^3(x-i0) =0$, 
namely, 
\beq\label{integral} 
\int dx\, \rho(x) F^2(x) = {\pi^2\over 3} \int dx\, \rho^3 (x)\,.
\eeq
Substituting (\ref{integral}) in (\ref{potential2}) we obtain the effective 
potential in its final form as 
\beqra\label{potential}
V &=&  {(\lambda_1-1)^2\over 8 m_1} \int dx {(\partial_{x} \rho_{1})^2\over 
\rho_{1}} + 
{(\lambda_2-1)^2\over 8 m_2} \int dx {(\partial_{x} \rho_{2})^2\over \rho_{2}}
\nonumber\\
&+& {\sqrt{\lambda_1}\over m_1}\int dx\, F(x)\, \partial_x \left({\lambda_1-1
\over 2}\rho_1 + 
{\lambda_2-1\over 2}\rho_2\right)\nonumber\\
&+& {\pi^2\sqrt{\lambda_1}\over 6m_1}\int dx\, \rho^3 (x)\nonumber\\
&+& \mu_{1} \left(N_{1} - \int dx \rho_{1}(x) \right) + \mu_{2} \left(N_{2} - 
\int dx \rho_{2}(x)\right)\,,
\eeqra
which can be shown easily to coincide with (\ref{vofh}).

\pagebreak

%


\newpage
\setcounter{equation}{0}
\setcounter{section}{0}
\renewcommand{\theequation}{B.\arabic{equation}}
\renewcommand{\thesection}{Appendix B:}
\section{Diagonalization of $\om_{ab}(k)$ in (\ref{omegaF})}
\vskip 5mm
\setcounter{section}{0}
\renewcommand{\thesection}{B}
The matrix $\om_{ab}(k)$ in (\ref{omegaF}) is of the general form 
\beq\label{form}
{\bf \om} = {\bf D} k^2 + \bv\bv^T |k|\,,
\eeq
where ${\bf D}$ is a diagonal matrix with diagonal entries 
$d_a = {1-\la_a\over 2m_a}$, and $\bv$ is an $F$-dimensional vector with 
components $v_a = cm_a\sqrt{{\pi\rho_{a,0}\over m_a}} = 
{\rm sign}\,(m_a)\sqrt{{\pi\rho_{a,0}\la_a\over m_a}}
\,,$ where we used (\ref{3body}) twice. After a standard computation we 
obtain the characteristic polynomial of (\ref{form}) as
\beq\label{characteristic}
P_F(z) \equiv \det (z - {\bf D} k^2 + \bv\bv^T |k|) = 
\det (z - {\bf D} k^2 )\left(1 - \bv^T{|k|\over z - {\bf D} k^2}\bv\right)\,.
\eeq
In the generic case, where all the diagonal elements $d_a$ of ${\bf D}$ are 
different from each other (and we shall henceforth assume this case), none 
of the eigenvalues of ${\bf \om}$ is an eigenvalue of ${\bf D}k^2$.
\footnote{This should be well-known. The reason for this is simple: the 
second factor in (\ref{characteristic}) has simple poles located at the 
eigenvalues of ${\bf D}k^2$, which cancel against zeros at the same points in 
the first factor in (\ref{characteristic}). If ${\bf D}$ is not degenerate, 
all those zeros are simple, and we end up with a nonvanishing result.} 
Therefore, all roots of $P_F(z)$ are obtained by finding the zeros of the 
second factor  in (\ref{characteristic}), namely, from 
\beq\label{roots}
\bv^T{|k|\over z - {\bf D} k^2}\bv = \sum_{a=1}^F\, {v_a^2 |k|\over 
z- d_a k^2} = 1 \,.
\eeq
It is clear from (\ref{form}) that all eigenvalues of ${\bf\om}$ 
vanish at least as fast as $|k|$ when the latter tends to zero. Alternatively, 
this can be seen by taking the limit $|k|\rightarrow 0$ in (\ref{roots}). 
Since all $v_a^2>0$, the roots must vanish at least like $|k|$. Thus, 
let us set $z = \zeta |k|$ in (\ref{roots}), and solve 
\beq\label{zeta}
\sum_{a=1}^F\, {v_a^2 \over \zeta- d_a |k|} = 1 
\eeq
for $\zeta$. In fact, we need to find $\zeta$ only to first order in $|k|$, 
as we are interested in the eigenvalues of (\ref{form}) only to order $k^2$. 

Let us first look for roots of (\ref{zeta}) which do not vanish as $|k|
\rightarrow 0$. We find only one such root, which at $|k|=0$ is given simply 
by $\zeta^{(0)} = \sum_a v_a^2 >0$.  
To find the leading correction to this root we expand (\ref{zeta}) in inverse 
powers of $\zeta$ to order $|k|$:
\beq\label{correction}
1- {1\over\zeta}\sum_{a=1}^F\,v_a^2 - {|k|\over\zeta^2}\,\sum_{a=1}^F\,
v_a^2 d_a + {\cO}(k^2) = 0\,.
\eeq
Only one root of this approximate quadratic equation converges to 
$\zeta^{(0)}\,,$ and to order $|k|$ it is given by 
\beq\label{zeta1}
\zeta^{(1)} = \sum_{a=1}^F\,v_a^2 + {\sum_{a=1}^F\,
v_a^2 d_a \over \sum_{a=1}^F\,v_a^2}\,|k|\,.
\eeq
Upon substituting the appropriate values of $v_a$ and $d_a$, (\ref{zeta1}) 
coincides with the expression (\ref{eigenvalues}) quoted in the text. In our
conventions $\lambda_a >0$ and ${\rm sign}\,m_a = {\sign}\,\rho_{a,0}\,.$ 
Therefore all $v_a$ are real, and consequently the coefficient 
$\sum_{a=1}^F\,v_a^2$ of $|k|$ in (\ref{eigenvalues}) is positive.

All other $F-1$ roots of (\ref{zeta}) vanish as $|k|$ when the latter 
tends to zero. To see this, restore the original roots $z = \zeta |k|$ 
and consider $\det (-{\bf \om}) = P_F(0) \sim (k^2)^{F-1}\,\left(k^2 + 
|k|\bv^T {\bf D}^{-1}\bv\right)\,.$ As $|k|\rightarrow 0$ this behaves
like $(k^2)^{F-1}\,|k|\,,$ proving our claim.

{\bf Acknowledgement}\\
This work was supported by the Ministry of Science and Technology of the Republic of Croatia under 
contract No. 0098003. JF wishes to thank the Center for Nonlinear Studies at the Los Alamos
National Laboratory, and in particular, Carl Bender and Bob Ecke, for the kind hospitality
extended to him during the completion of this work.


\end{document}